\documentclass[useAMS,usenatbib]{mn2e}
\usepackage{graphicx}
\def\aj{AJ}%
\def\araa{ARA\&A}%
\def\apj{ApJ}%
\def\apjl{ApJ}%
\def\apjs{ApJS}%
\def\aap{A\&A}%
\def\aapr{A\&A~Rev.}%
\def\mnras{MNRAS}%
\usepackage{times}
\usepackage{amsmath}
\usepackage{amssymb}
\usepackage{color}
\usepackage{subfigure}

\usepackage{multirow}

\title[The SOAR Gravitational Arc Survey]{{The SOAR Gravitational Arc Survey -- I. Survey overview and photometric catalogues\thanks{Based on observations obtained at the Southern Astrophysical Research (SOAR) telescope, which is a joint project of the Minist\'{e}rio da Ci\^{e}ncia, Tecnologia e Inova\c{c}\~{a}o (MCTI) da Rep\'{u}blica Federativa do Brasil, the U.S. National Optical Astronomy Observatory (NOAO), the University of North Carolina at Chapel Hill (UNC) and Michigan State University (MSU).}}}

\author[Furlanetto et al.]{Cristina Furlanetto$^{1,2}$\thanks{E-mail:cristina.furlanetto@ufrgs.br}, Bas\'ilio X. Santiago$^{1,2}$, Mart\'in Makler$^{2,3}$, Eduardo S. Cypriano$^{2,4}$,
\newauthor Gabriel B. Caminha$^{2,3}$, Maria Elidaiana da Silva Pereira$^{2,3}$, Angelo Fausti Neto$^{2}$,
\newauthor Juan Estrada$^{5}$, Huan Lin$^{5}$, Jiangang Hao$^{5}$, Timothy A. McKay$^{6}$,
\newauthor Luiz Nicolaci da Costa$^{2,7}$ and Marcio A. G. Maia$^{2,7}$\\
$^{1}$Departamento de Astronomia, Universidade Federal do Rio Grande do Sul, Av. Bento Gon\c{c}alves 9500, Porto Alegre, RS 91501-970, Brazil\\ 
$^{2}$Laborat\'orio Interinstitucional de e-Astronomia, Rua Gen. Jos\'e Cristino 77, Rio de Janeiro, RJ 20921-400, Brazil\\ 
$^{3}$Centro Brasileiro de Pesquisas F\'isicas, Rua Dr. Xavier Sigaud 150, Rio de Janeiro, RJ 22290-180, Brazil\\ 
$^{4}$Instituto de Astronomia, Geof\'isica e Ci\^encias Atmosf\'ericas, Universidade de S\~ao Paulo, Rua do Mat\~ao 1226, S\~ao Paulo, SP 05508-090, Brazil \\ 
$^{5}$Center for Particle Astrophysics, Fermi National Accelerator Laboratory, Batavia, IL 60510, USA\\ 
$^{6}$Department of Physics, University of Michigan, Ann Arbor, MI 48109, USA\\ 
$^{7}$Observat\'orio Nacional, Rua Gen. Jos\'e Cristino 77, Rio de Janeiro, RJ 20921-400, Brazil 
}

\begin{document}

\date{Accepted 2013 February 27. Received 2013 February 25; in original form 2012 July 4}

\pagerange{\pageref{firstpage}--\pageref{lastpage}} \pubyear{2013}

\maketitle

\label{firstpage}

\begin{abstract}
\\
We present the first results of the SOAR (Southern Astrophysical Research) Gravitational Arc Survey (SOGRAS). The survey imaged 47 clusters in two 
redshift intervals 
centered at $z=0.27$ and $z=0.55$, targeting the richest clusters in each interval. Images were obtained in the $g'$, $r'$ and $i'$ bands 
using the SOAR Optical Imager (SOI), with a median seeing of $0.83$, $0.76$ and $0.71$ arcsec, respectively, in these filters. Most of the survey 
clusters are located within the Sloan Digital Sky Survey (SDSS) Stripe 82 region and all of them are in the SDSS footprint. Photometric calibration 
was therefore performed using SDSS stars located in our SOI fields. 
We reached for galaxies in all fields the detection limits of $g \sim 23.5$, $r \sim 23$ and $i \sim 22.5$ for
a signal-to-noise ratio (S/N) = 3. As a by-product
of the image processing, we generated a source catalogue with 19760 entries, the vast majority of which are galaxies, where we list their 
positions, magnitudes and shape parameters. 
We compared our galaxy shape measurements to those of local galaxies and concluded that they were not strongly affected by seeing. From the catalogue
 data, we are able to identify a red sequence of galaxies in most clusters in the lower $z$ range. 
We found 16 gravitational arc candidates 
around 8 clusters in our sample. They tend to be bluer than the central 
galaxies in the lensing 
cluster. A preliminary analysis indicates that $\sim 10\%$ of the clusters have arcs around them, with a possible indication of a larger efficiency 
associated to the high-$z$ systems when compared to the low-$z$ ones. 
Deeper follow-up images with Gemini strengthen the case for the strong lensing nature of the candidates found in this survey. 

\end{abstract}

\begin{keywords}
gravitational lensing: strong -- surveys -- galaxies: clusters: general
\end{keywords}

\section{Introduction}

Tracing the evolution of galaxy cluster properties, in particular their mass distribution, has important implications for their use as cosmological 
probes, for understanding the nature of dark matter and dark energy and to 
constrain galaxy evolution. A unique way to assess the mass distribution in clusters is through the arcs produced by strong gravitational 
lensing \citep{BN92,Hattori99, Guzik2002, Mandelbaum2006, Treu10, KN12}. In particular, the statistics of gravitational arcs may provide constraints 
on cosmological parameters and on scenarios of structure formation \citep{Bartelmann98,Bartelmann2003, Golse2002, Meneghetti2004, Kochanek2006, Hilbert2007, Vuissoz2007,Zieser2012}.

This motivated arc searches to be conducted, both in images from wide field surveys \citep{Gladders03, sdss1,legacy,CASSOWARYmethod, SBAS7,CS82, RCS2,more11,WenHanJiang2011,Bayliss2012,Wiesner2012}, as well as in fields targeting known clusters, with observations from the ground  \citep{Luppino99,Zaritsky03,Hennawi2008,kausch2010} and from space \citep{Smith2005, Sand2005, Horesh2010}. 
Upcoming wide field imaging surveys, such as the Dark Energy Survey\footnote{\texttt{www.http://www.darkenergysurvey.org/}} \citep[DES;][]{annis05,des05}, which started operations this year, 
are expected to detect about an order of magnitude more arcs than the current largest surveys.

For over a decade there has been a debate about the compatibility of the observed arc abundance with  theoretical modeling. 
\citet{Bartelmann98} suggested an apparent overabundance by approximately an order of magnitude of giant arcs on the sky as 
compared to $\Lambda$CDM predictions. Subsequent comparison works with limited statistics confirmed that the number of giant arcs on 
the sky is underpredicted by the $\Lambda$CDM cosmological model \citep{Luppino99, Gladders03, Zaritsky03, Li06}. 

However, more recent studies including several factors that were not present in the first predictions, such as using simulations at the image level  \citep{Horesh11, Boldrin2012}
and including mergers \citep{Redlich2012} have reduced the overall discrepancy between observed and 
predicted arc abundances \citep{Dalal04, Horesh05, Hennawi07}.
In particular, \citet{Horesh11} have carried out a study using simulations,  
where gravitational arcs are generated from ray-tracing of realistic sources 
from the Hubble Ultra Deep Field through clusters from the Millennium  $N$-body simulation. The fraction of arcs per cluster in 
the simulated samples is compared to an arc sample in clusters with similar properties, using the same methods to identify the arcs 
in both samples. They find an overall consistency of the observed and simulated samples, at least in the redshift interval $0.3<z<0.6$.
On the other hand, another comparison between arcs in x-ray selected clusters on data and on simulations \citep{MeneghettiXselected2011}, still found a disagreement among them, although the discrepancy is smaller than the earlier estimates a decade ago. Also, including several baryonic effects on the simulations does not solve the remaining discrepancy \citep{Killedar2012}.

If on the one hand the arcs statistics problem may have been solved or at least mitigated, issues remain regarding the 
variation of arc abundance with respect to the cluster redshift. For example, \citet{Gladders03} found an over-abundance of 
arcs in high-redshift clusters as compared to lower redshift ones. \citet{Gonzalez12} found arcs in a cluster at $z = 1.75$, 
which should not be present at their image depths according to their modeling. \citet{Horesh11} found an under-prediction of clusters 
at $ z \sim 0.2$ as compared to observations analyzed in  \citet{Horesh05}. These discrepancies could be due to the evolution of 
cluster structure with redshift (including the role of baryons) and/or to selection effects of the samples (e.g., X-ray versus optical selection). Caminha et al. (in preparation) 
model the 
variation of arc abundance with lens redshift stressing the effect of magnification on the expected distribution and finding an 
increase of arc incidence with $z$.

The main motivation for the {\it SOAR Gravitational Arc Survey} (SOGRAS) was to constrain the variation of strong lensing 
efficiency as a function of cluster redshift, comparing the results with theoretical expectations. For this sake, we have 
designed a survey targeting clusters distributed in two redshift bins centered at $z \sim 0.3$ and $z \sim 0.5$. A total 
of 47 clusters were imaged in the $g'$, $r'$ and $i'$ bands with the 4.1 Southern Astrophysical Research Telescope (SOAR) 
from mid 2008 to early 2011.

The arcs and other strong lensing features found can be used to constrain the individual masses of the 
clusters \citep[e.g.,][]{Cypriano2005}. Another valuable information that can be drawn from the data is an estimate of 
ensemble cluster masses in each $z$ bin with weak lensing, by stacking the profile of the scaled tangential distortion of 
background sources of all clusters in that bin. This technique has been applied successfully 
\citep{Sheldon2001, Sheldon2004, Johnston2007} and leads to an averaged overall mass for the clusters.  

Another motivation was to use this dataset as a test bed for tools being developed for DES, in particular for gravitational 
arc studies, including testing automated arc-finders (using either morphology or colour) and methods to measure arcs 
properties \citep[e.g.][]{Furlanetto12b}. Indeed SOGRAS has comparable depth and seeing conditions as expected from DES 
and covers 3 of the 5 DES bands.

Finally, SOGRAS can be seen as a pathfinder for a high resolution arc survey with SOAR using the recently commissioned SOAR 
Adaptive Module\footnote{\texttt{www.ctio.noao.edu/new/Telescopes/SOAR/Instruments/SAM/}} \citep{SAM08,SAM10}.

In this paper we report on the overall properties of the survey, from the target selection and observations to data reduction 
and photometric calibration. We present the photometric catalogs and discuss new gravitational arc candidates found by visual 
inspection. Detailed results on arc analyses and comparison with theoretical modeling will be presented in a accompanying paper.

As a by-product of the survey a large catalogue of galaxies in the cluster fields was generated, with astrometric, photometric, 
and morphological information. This catalogue was used to separate cluster members (through the red sequence in the colour-magnitude 
diagrams) from field galaxies, and will be useful for future analysis of galaxy evolution. 

The outline of this paper is as follows: \S \ref{survey} describes the survey, including information on sample selection, 
observational and image details. In \S \ref{data_reduction} we describe the data reduction, including astrometric 
and photometric calibrations. We also carefully assess the quality of our photometry. The resulting galaxy catalogue is 
presented in \S \ref{catalogs}. In the same section we present the first sample of arc candidates. Finally, 
in \S \ref{conclusion} we present our summary and closing remarks.

\section{The survey}
\label{survey}

We have designed the survey to image a sample of galaxy clusters, equally split into two disconnected redshift bins, 
one at $0.20<~z_{phot}<0.35$ (the ``low-$z$'' bin) and the other at $0.50<~z_{phot}<0.60$ (the ``high-$z$'' bin), to have a 
``leverage arm'' to constrain the evolution of arc incidence between these two intervals.

The low-$z$ bin was chosen such that there are enough reasonably rich clusters in this bin on the survey footprint 
(see \S \ref{footprint82}) and to avoid having a too small arc probability. The high-$z$ bin was determined by the 
availability of optical cluster catalogs in the survey footprint and by the requirement of having enough background 
sources to allow for a weak lensing analysis by stacking the clusters in this bin (to have an overall estimate of the 
cluster masses).

All clusters were observed with the SOAR telescope, located on Cerro Pach\'on in the Chilean Andes, with the SOAR Optical 
Imager (SOI). The choice of telescope and instrument is motivated by the typical site seeing ($\simeq 0.8''$) and 
detector pixel size, which yield the required image quality for gravitational arc detection\footnote{As is well known, 
the detectability of gravitational arcs is very sensitive to the PSF FWHM, because the seeing tends to decrease their 
length-to-width ratios and dilutes their surface brightness \citep[e.g.][]{Cypriano01}.}. 
 
The observations of all our targets were carried out in queue-scheduled mode assuring that our quality 
requirements were met. Therefore, this survey provides a fairly homogeneous sample, in the sense that all images 
were obtained in similar conditions, with the same instrument and filters and same exposure time, and is therefore 
well suited for a comparison of arc incidence.  
 
The exposure time was determined by a balance between the number of clusters to be observed and the depth achieved 
for each cluster field for a given total observing time. Using the model for the number of arcs expected per cluster 
(as a function of limiting magnitude, cluster redshift, etc.) given in Caminha et al. (in preparation) 
and the exposure time 
calculator\footnote{\texttt{http://www.noao.edu/gateway/ccdtime/}}, we found that the maximum total number of arcs is reached for integration times of about 10 min (in one single filter).

We also required to image in 3 bands such that colour information could be gathered, since this is an important 
information for discriminating gravitational arcs from cluster tidal features and for identifying multiple images. 
Furthermore, the colour information helps mitigating the contamination by foreground objects for weak-lensing mass reconstructions.

\subsection{SDSS and Stripe 82}
\label{footprint82}

The baseline footprint for the targeting of our survey was the SDSS ``Stripe 82'', a $275 \deg^2$ equatorial stripe 
(over $-50<RA<59$; $-1.25<DEC<1.25$), which was scanned multiple times in the fall seasons of 2000--2007 as part of a 
supernovae search, leading to a much deeper survey. The final Stripe 82 coadded data (hereafter {\it coadd}) 
reaches $r\sim 23.5$ for galaxies, i.e. 2 mag fainter than the main SDSS survey \citep{Annis11}.

The availability of these data allowed for the construction of deeper cluster catalogs, well suited for our target selection. 
In particular, red-sequence based cluster catalogs started to be produced as the first coadds of several stripe 82 visits were 
created. While single pass SDSS cluster catalogs reached up to $z \sim 0.3$ \citep{maxbcgK}, catalogs obtained from 
the {\it coadd} reach $z \sim 0.6$, matching our requirement for the high-$z$ bin, having at the same time some leverage 
arm with respect to the low-$z$ bin and still allowing for a stacked weak lensing measurement from our data.

At present, besides the deeper SDSS imaging, parts of Stripe 82 have been covered by a wealth of multi-wavelength data, 
such as the UKIDSS Large Area Survey in the {\it YJHK} bands \citep{UKIDSS}, deep {\it GALEX} UV imaging \citep{Martin05}, the SHELA \citep{SpitzerProposal} and SpIES \citep{SpIESProposal} surveys with 
Spitzer/IRAC, and the HeLMS 
\citep{HerMES} and
SPIRE \citep{SPIREaph} surveys with Herschel. 
At longer wavelengths, the
whole Stripe lies within the footprint of Atacama Cosmology Telescope equatorial survey \citep{sehgal12} and 80 deg$^2$ of the Stripe have deep VLA data \citep{Hodge11}. 
Stripe 82 has also a very high density of spectroscopic redshifts, with redshift measurements from SDSS \citep{dr7}, 2dF \citep{2dFGRS, 2QZ}, 
2SLAQ \citep{2SLAQ}, 6dF \citep{6dF}, 
DEEP2 \citep{DEEP2}, VVDS 
\citep{Garilli08}, PRIMUS \citep{PRIMUS}, SDSS-III/BOSS
\citep{BOSS12} and WiggleZ \citep{wigglez}. 
This region is thus emerging as a ``deep 
extragalactic survey field'', a precursor to DES and LSST, with an impressive array of multi-wavelength observations already in hand 
or in progress. SOGRAS may be used as a test case for combining the good imaging data from SOAR with this large set of complementary 
data, which strengthens the case to carry out most of our selection in this field.

In particular, during the second semester of 2010 until early 2011, 170 deg$^2$ of Stripe 82 were imaged in the $i$ band 
for the CFHT/Megacam Stripe 82 Survey \citep[CS82;][Erben et al., in preparation]{CS82}, 
providing data down to $i=23.5$ obtained in excellent 
seeing conditions (median seeing of $0.6''$), enabling precision weak lensing measurements. This dataset is particularly 
synergistic with SOGRAS. CS82 data will allow us to study the clusters imaged for SOGRAS at much larger radii and provide 
weak lensing measurements for them. On the other hand the SOGRAS data is useful for quality assessment on this new dataset 
around clusters. For example, the star-galaxy separation and the determination of background and foreground sources can be 
tested in these fields thanks to the colour information at higher depths and better seeing than the SDSS photometric data.

\subsection{Target selection and final galaxy cluster sample} 

The survey was carried out in two seasons, the first during semester 2008B and the second in 2010B \citep{sogras08,sogras10}. 
Since the cluster catalogs and status of the Stripe 82 coadd evolved during the two seasons, different catalogs were used for 
the selection. The procedure was nevertheless the same for both seasons: selecting the richest cluster catalogs in the same two 
redshift bins and requiring the same imaging conditions and instrument configurations. Therefore the two sets of observations 
are considered as a single dataset.

The cluster selection for the 2008B season was carried out using a combination of three unpublished cluster catalogs on 
Stripe 82 (J. Hao, T. McKay, et al.). The cluster finding methods were based on the red-sequence, accounting for its 
variation with redshift, and are precursors of the {\it Gaussian Mixture Brightest Cluster Galaxy} (GMBCG) cluster 
finder \citep{Hao10}. However they differ in their likelihoods and the radial profiles used.
All were run in the coadded data available in late 2006, providing an estimate of the cluster photometric 
redshift ($z_{phot}$) and richness\footnote{Roughly the number of red-sequence galaxies in the cluster with luminosity above $L^\star/2$, where $L^\star$ is the characteristic luminosity in the \citet{Schechter76} luminosity function.}. We selected the richest clusters from these catalogs in the two redshift bins and ranked them by richness. The centres of the pointings were chosen as the cluster centre, defined as the position of the Brightest Cluster Galaxy (BCG) as determined by the cluster finding method.

We complemented this {\it main sample}, with an {\it extra sample} consisting of clusters detected on SDSS Data Release (DR) 6 
data \citep{SDSSDR6} --- not necessarily on stripe 82 --- from the MaxBCG \citep{maxbcgK} catalogue that matched 
ROSAT x-ray sources and had good observability from SOAR on that semester. No redshift restriction was applied 
to this sample, which was chosen to improve our chances of finding arc systems and for scheduling flexibility 
(i.e. to allow observations to be made when observing conditions were not suitable for Stripe 82). Naturally, 
clusters observed from this sample are not suitable for our arc statistics purposes.

A visual inspection of SDSS single pass images using the catalogue Archive Server\footnote{\texttt{http://cas.sdss.org/dr6/en/tools/chart/list.asp}} \citep{Thakar2008, dr7} was made in order to avoid clusters close to bright stars, which could jeopardize the observations. We discarded all clusters that show diffraction spikes and star halos within a $\sim 6.5' \times 6.5'$ field around the cluster center. We also discarded fields with saturated stars within $3'$ from the cluster center, imposing stronger limits on the magnitude closer to the center (e.g. $mag \lesssim 14$ for $\theta \lesssim 1'$). This eliminated $\sim 25\%$ of the selected fields.
Clusters that appeared to have more than a single structure (e.g. could be line-of-sight superpositions)
from this visual inspection were also avoided, eliminating $\sim 10\%$ of the selected clusters. While performing the visual inspection, 
we ignored any potential arc feature to avoid biasing the sample. The final result of this process was a set of two lists (one for each redshift bin) ordered by richness, containing a total of 60 selected clusters. The observers were told to select randomly among these lists, choosing the 
highest ranked object for which the observing conditions were favorable. 

A total of 18 fields were observed in that season, 13 corresponding to clusters selected in the the high-$z$ bin, 4 in the low-$z$ bin and 1 from the extra sample. 

For the 2010B season we made a new selection of targets using the detections
from a GMBCG cluster catalogue constructed using the complete {\it coadd} data \citep{Annis11}. To exploit 
the synergy with CS82, only clusters in the footprint of this survey were selected. Again we selected the clusters on the two redshift bins and kept the richest ones. As in the 2008 sample, we also included an extra sample 
with the same objects selected for that season. The visual selection procedure was the same as for the 2008 season.

Initially, 26 fields were observed, corresponding to 12 clusters detected in the high-$z$ bin, 11 in the low-$z$ bin and 3 from the extra sample.\footnote{Two fields had a large overlap with pointings from the 2008B season. Since the possible arc candidates will be located close to the cluster center, we considered those fields as a single one.
In this case we kept only the 2010B fields in our imaging sample, both due to the better quality of the imaging, as well as to the improved determination of cluster properties from the catalogue used for this selection. This improved cluster detection made the object observed as part of the extra sample in 2008B to correspond to a pointing in the low-z sample.\label{overlap}}
At that point, the SOGRAS program had still telescope time allocated, but the observability of Stripe 82 was unfavorable. Since there were more clusters observed from the high-$z$ bin, than in the low-$z$ one, clusters could be selected only in the later, thus requiring only shallower imaging. Therefore an {\it auxiliary sample} was chosen, following the same selection criteria as the low-$z$ one, but choosing clusters at higher $RA$ in an equatorial region  covered by SDSS single pass imaging. These clusters were taken from a GMBCG catalogue based on 
SDSS DR7 data \citep{Hao10}. 
Seven fields from this sample
were observed, completing the survey. Two of these had large overlaps on their central regions with other pointings on the same sample. Therefore we consider the auxiliary sample as composed by only 5 independent fields.

Thus a total of 47 independent cluster fields were observed for this project with SOAR (accounting for the overlapping fields in the sample).

\begin{figure*}
\centering
{
\includegraphics[scale=0.5]{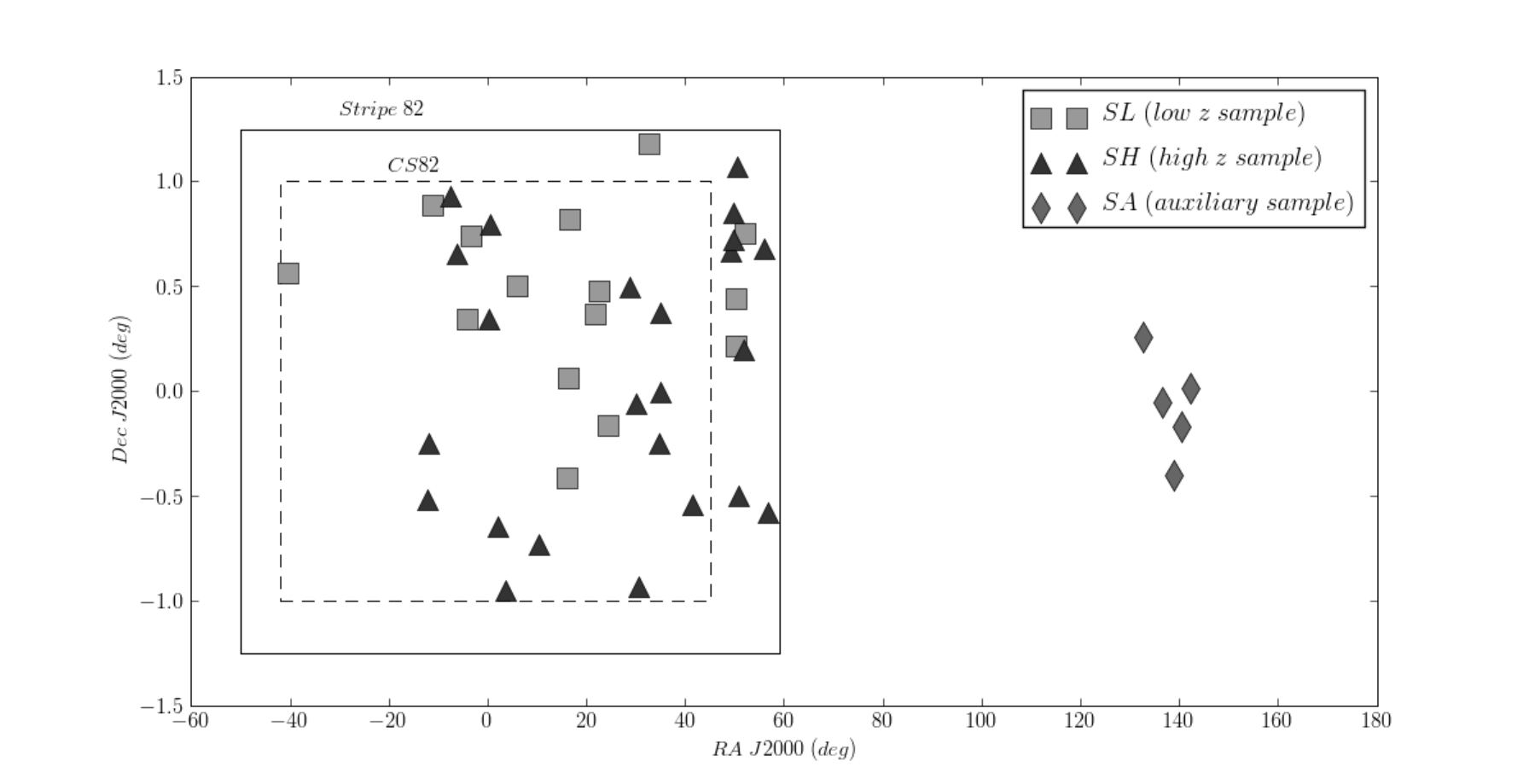}

\caption{On-sky distribution of SOGRAS clusters. The area bounded by the dashed line is the CS82 footprint, which essencially is a subarea of the Stripe 82 footprint (bounded by the solid line). The auxiliary sample is outside of Stripe 82, but is still in the main SDSS footprint. For scale reasons, the 3 clusters of the extra sample whose positions are far from the equatorial stripe (SOGRAS0940+0744, SOGRAS1023+0413 and SOGRAS1054+1439) are not shown. }
\label{footprint}
}
\end{figure*}

\begin{figure}
\centering
\includegraphics[scale=0.42]{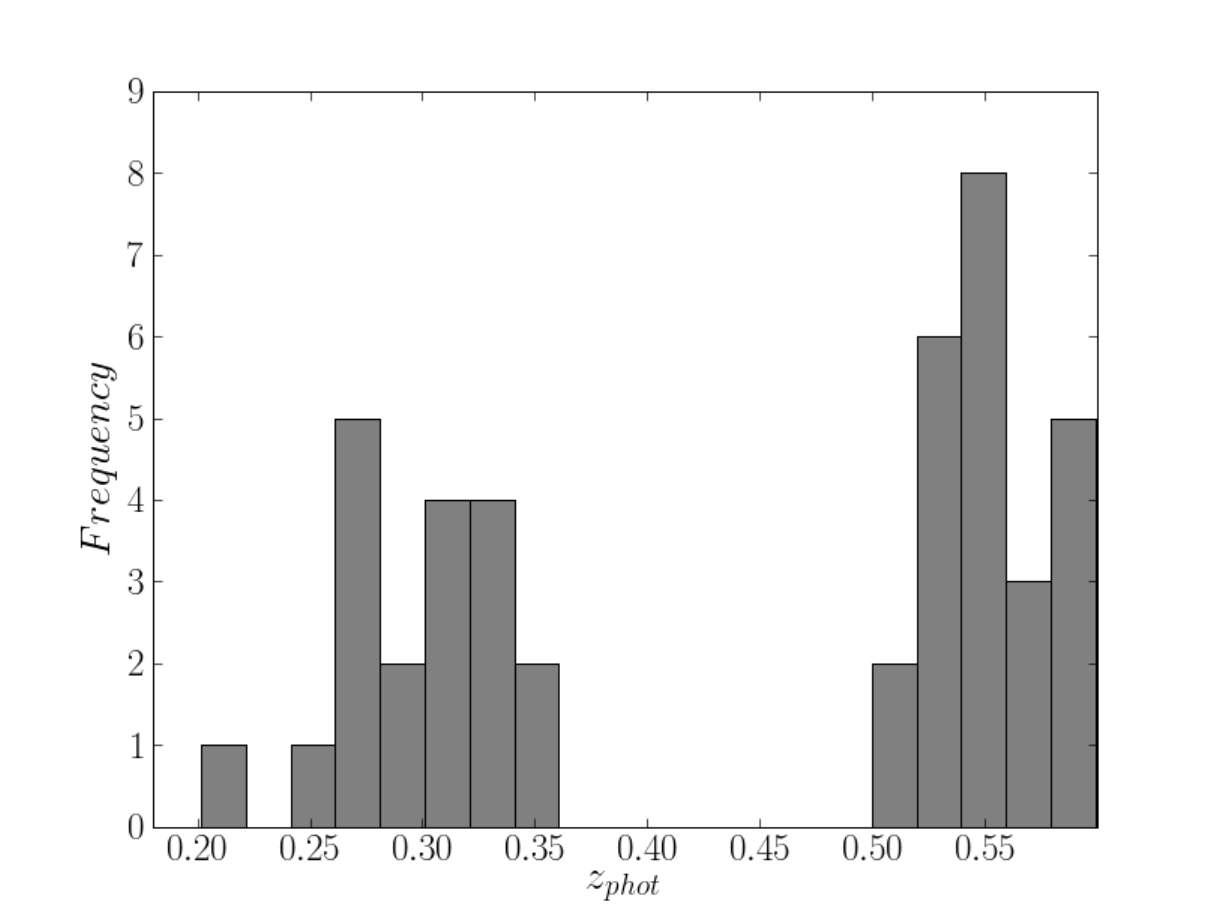}
\caption{Distribution of main sample clusters as a function of photometric redshift.}
\label{z_n_sogras}
\end{figure}

The on-sky distribution of the observed fields is shown in Figure \ref{footprint}. From the 47 observed clusters, 39 are in the main sample, 5 are in the auxiliary 2010 sample and 3 are in the extra sample. For arc statistics analysis, the auxiliary sample can be added to the main sample, since its clusters were selected following the same criteria for redshift and richness. Although the 
clusters in the auxiliary sample were taken from a shallower photometric catalog, the same cluster finder algorithm (GMBCG) was used to generate the cluster catalogue in the main SDSS area and in the {\it coadd} 
and it is known to be complete for the low-$z$ clusters. The resulting split in $z$ was 24 clusters in the high-$z$ bin and 20 in the low-$z$ bin. The 3 clusters in the extra sample cannot be included in a statistical analysis because they were selected following other criteria. The distribution of photometric redshifts for the SOGRAS clusters is shown in Figure \ref{z_n_sogras}. 

A summary of the properties of the SOGRAS clusters is given in Table \ref{targets_table}. The mean photo-z (fourth column) uncertainty is 0.03 \citep{Reis2012}. The fifth column,  $N_{\rm gals}^{i}$, shows the richness as taken from the original selection catalogs. As mentioned above, these catalogs were obtained from different methods (three for 2008B, one for 2010B) and different data (different number of coadds for 2008B and 2010B, plus the single pass data for the Auxiliary Sample). Therefore, those richnesses cannot be compared directly. To provide a more uniform estimate of the richness, we have run a single code, the Error Corrected Gaussian Mixture Model \citep[ECGMM,][]{Hao10}, 
over all observed fields on the final Stripe 82 coadd data, providing a new richness estimate for these clusters (sixth column, $N_{\rm gals}^{\rm GM}$). We used the run on the DR7 data  \citep{Hao10} to provide a richness for the objects outside Stripe 82. However, as the single pass and coadd data have different $S/N$, again the richness cannot be compared. Using the richness from both the single pass, $N_{\rm gals}^{\rm DR7}$, and coadd data, $N_{\rm gals}^{\rm coadd}$, for clusters on Stripe 82 we obtained a mean relation connecting them: $N_{\rm gals}^{\rm coadd} = 0.40 \times N_{\rm gals}^{\rm DR7}  + 12.3$.
Applying this relation to the objects on our sample outside Stripe 82 we obtain a ``renormalized'' richness estimate\footnote{These values should be interpreted with care, since it is well known that the richness from optical clusters have a very large scatter and the relation above is only a mean relation.} that should be more comparable among all clusters in the sample ($N_{\rm gals}^{\rm GM}$). 

\begin{table*}
\begin{minipage}{115mm}
\caption{Summary of the SOGRAS galaxy cluster sample.}
\label{targets_table}
\begin{tabular}{l c c c c c c} 
\hline
Cluster ID & RA  & Dec  &  $z_{phot}$ & $N_{\rm gals}^{i \,a}$ & $N_{\rm gals}^{\rm GM\,b}$ & Observation date \\ 
  & (J2000) & (J2000) & &  \\
\hline

SOGRAS0001+0020				&	00:01:01	&	00:20:17	&	0.538		&	39					&	6	&	2008-11-20	\\	
SOGRAS0001+0047				&	00:01:56	&	00:47:21	&	0.527		&	37					&	6	&	2008-10-03	\\	
SOGRAS0008-0038				&	00:08:21	&	-00:38:45	&	0.523		&	32					&	53	&	2010-12-02	\\	
SOGRAS0014-0057				&	00:14:54	&	-00:57:08	&	0.535		&	62					&	46	&	2010-10-31	\\	
SOGRAS0024+0030				&	00:24:00	&	00:30:07	&	0.292		&	33					&	30	&	2010-11-01	\\	
SOGRAS0041-0043				&	00:41:09	&	-00:43:49	&	0.564		&	28					&	26	&	2010-12-02	\\	
SOGRAS0104-0024				&	01:04:24	&	-00:24:51	&	0.266		&	26					&	24	&	2010-12-08	\\	
SOGRAS0104+0003				&	01:04:55	&	00:03:36	&	0.272		&	85					&	66	&	2010-10-31	\\	
SOGRAS0106+0049				&	01:06:07	&	00:49:10	&	0.263		&	35					&	32	&	2010-10-31	\\	
SOGRAS0127+0022				&	01:27:13	&	00:22:06	&	0.338		&	26					&	40	&	2010-11-01	\\	
SOGRAS0130+0028				&	01:30:36	&	00:28:39	&	0.335		&	24					&	27	&	2010-12-08	\\	
SOGRAS0137-0009				&	01:37:29	&	-00:09:56	&	0.341		&	37					&	38	&	2010-10-31	\\	
SOGRAS0155+0029				&	01:55:38	&	00:29:42	&	0.525		&	36					&	9	&	2008-11-20	\\	
SOGRAS0200-0003				&	02:00:33	&	-00:03:46	&	0.580		&	41					&	47	&	2010-10-31	\\	
SOGRAS0202-0055				&	02:02:23	&	-00:55:57	&	0.599		&	43					&	34	&	2010-10-31	\\	
SOGRAS0210+0110				&	02:10:56	&	01:10:44	&	0.276		&	88					&	35	&	2008-11-21	\\	
SOGRAS0218-0014				&	02:18:45	&	-00:14:52	&	0.502		&	34					&	73	&	2010-11-01	\\	
SOGRAS0219+0022				&	02:19:49	&	00:22:25	&	0.531		&	36					&	35	&	2008-10-03	\\	
SOGRAS0220-0000				&	02:20:03	&	-00:00:18	&	0.555		&	28					&	42	&	2010-12-08	\\	
SOGRAS0245-0032				&	02:45:27	&	-00:32:36	&	0.580		&	54					&	42	&	2010-10-31	\\	
SOGRAS0316+0039				&	03:16:46	&	00:39:54	&	0.554		&	31					&	7	&	2008-11-20	\\	
SOGRAS0319+0042				&	03:19:25	&	00:42:52	&	0.546		&	32					&	4	&	2008-11-04	\\	
SOGRAS0319+0050				&	03:19:44	&	00:50:55	&	0.576		&	40					&	23	&	2008-11-20	\\	
SOGRAS0320+0012				&	03:20:47	&	00:12:43	&	0.255		&	24					&	9	&	2008-11-21	\\	
SOGRAS0321+0026				&	03:21:11	&	00:26:20	&	0.309		&	47					&	34	&	2008-11-21	\\	
SOGRAS0321+0103				&	03:21:57	&	01:03:59	&	0.549		&	31					&	2	&	2008-11-04	\\	
SOGRAS0322-0030				&	03:22:56	&	-00:30:06	&	0.543		&	41					&	30	&	2008-11-21	\\	
SOGRAS0327+0011				&	03:27:09	&	00:11:32	&	0.549		&	31					&	27	&	2009-01-02	\\	
SOGRAS0328+0044				&	03:28:15	&	00:44:51	&	0.322		&	41					&	30	&	2009-01-02	\\	
SOGRAS0343+0041				&	03:43:57	&	00:41:31	&	0.511		&	33					&	0	&	2008-11-04	\\	
SOGRAS0346-0035				&	03:46:39	&	-00:35:03	&	0.541		&	31					&	10	&	2009-01-02	\\	
SOGRAS0850+0015$^{c,d}$		&	08:50:23	&	00:15:36	&	0.202		&	42					&	29	&	2011-01-11	\\	
SOGRAS0905-0003$^c$			&	09:05:52	&	-00:03:19	&	0.305		&	30					&	24	&	2011-01-12	\\	
SOGRAS0916-0024$^{c,d}$		&	09:16:09	&	-00:24:16	&	0.345		&	78					&	43	&	2011-01-11	\\	
SOGRAS0921-0010$^c$			&	09:21:41	&	-00:10:18	&	0.305		&	35					&	26	&	2011-01-12	\\	
SOGRAS0928+0000$^c$			&	09:28:45	&	00:00:55	&	0.307		&	47					&	31	&	2011-01-12	\\	
SOGRAS0940+0744$^e$			&	09:40:53	&	07:44:25	&	0.390		&	68					&	39	&	2011-01-11	\\	
SOGRAS1023+0413$^e$			&	10:23:39	&	04:13:08	&	0.465		&	42					&	29	&	2011-01-11	\\	
SOGRAS1054+1439$^e$			&	10:54:17	&	14:39:04	&	0.328		&	118					&	59	&	2011-01-12	\\	
SOGRAS2118+0033				&	21:18:49	&	00:33:37	&	0.276		&	68					&	53	&	2010-10-31	\\	
SOGRAS2311-0030				&	23:11:06	&	-00:30:59	&	0.594		&	34					&	39	&	2010-11-01	\\	
SOGRAS2312-0015				&	23:12:52	&	-00:15:02	&	0.588		&	51					&	40	&	2010-10-03	\\	
SOGRAS2315+0053				&	23:15:45	&	00:53:12	&	0.326		&	32					&	37	&	2010-11-01	\\	
SOGRAS2330+0055				&	23:30:09	&	00:55:51	&	0.548		&	40					&	40	&	2010-11-01	\\	
SOGRAS2335+0039$^d$			&	23:35:42	&	00:39:20	&	0.564		&	46					&	23	&	2010-11-01	\\	
SOGRAS2343+0020$^{d}$				&	23:43:34	&	00:20:37	&	0.269		&	55					&	37	&	2010-10-31	\\	
SOGRAS2346+0044				&	23:46:30	&	00:44:23	&	0.291		&	37					&	28	&	2010-11-01	\\	
\hline
\end{tabular}
$^a$ Richness taken from the original selection catalogs.\\
$^b$ Richness obtained with the ECGMM method on the final stripe-82 coadded data. In the case of clusters from the auxiliary data, this value is converted from the single pass data as described in the text.\\
$^c$ Cluster from the auxiliary sample.\\
$^d$ These fields have overlapping observations.\\ 
$^e$ Cluster from the extra sample.\\
\end{minipage}
\end{table*}

\subsection{Observations}

Observations of all our targets were carried out with SOI, which consists of a mini-mosaic of two E2V CCDs, each one with 
$4096 \times 2048$ pixels, covering a field of view of 5.25' $\times$ 5.25'. A 2 $\times$ 2 binning was used, yielding a 
detector scale of $0.154''$/pixel. The exposures were taken in fast read-out mode. Bias and flat-field images were also 
observed on the same nights, except for the nights 2008-10-03 and 2008-11-21. 

Each target field in our programme was imaged in the $g'$, $r'$ and $i'$ filters. For each filter we had three exposures of 
180 sec, which were slightly dithered by $\sim 10''$ in the direction perpendicular to the gap between the SOI CCDs. This 
dithering pattern allowed us to fill the gap region as well as to remove CCDs defects and cosmic ray hits. The centre of 
the target cluster was placed at $30''$ from the gap, to the east direction. 

To achieve the image quality needed for the survey in order to increase the arc detection efficiency, 
we required a seeing FWHM smaller than $0.8''$. We also required an airmass constraint of $X \le 1.5$, which is adequate for targets close to the celestial equator imaged from SOAR. Finally, we required nights with $\leq 7$ days from New Moon,
to reduce the noise over the images, mainly in the $g'$ band. 

\section{Data reduction}
\label{data_reduction}

The individual exposures of each SOGRAS field were bias-subtracted and flat-fielded using standard tasks from the MSCRED package of the Image Reduction and Analysis Facility (IRAF). For the nights in which there were no bias and flat-field exposures, exposures of the previous night were used. 

Custom codes were employed to remove the fringe pattern from $i'$ band images, since the defringing performed by IRAF MSCRED tasks was not satisfactory and we found that it could be improved upon. Our codes identify pixels on top of the fringing pattern and compute median counts for them, both for
the sky subtracted science and fringing correction images, computing the ratio between the two. They then scale the correction frame by this ratio before subtracting the pattern from the science image. The fringe amplitude (i.e.,
the typical difference between peak and valley in the fringe pattern) 
in the raw data was approximately $4\%$ of the sky and was reduced to levels smaller than $1\%$ of the sky by our defringing method for most of our $i'$ band images. However, we noticed a remaining fringing residual, especially in the west CCD chip, for about $12\%$ of our $i'$ band images. 
This residual fringing amplitude was at worst $\sim 2\%$ of the sky level. 
We also noticed a small difference in the counts of the four SOI amplifiers ($< 2\%$) and a difference in the noise level between the two CCDs. 

The original Multiple Extension Format (MEF) files were then converted into Flexible Image Transport System (FITS) files using the task {\it soimosaic} from the SOAR/SOI IRAF package. 

Exposures of the same filter of each target were aligned and combined into a stacked image, by taking the median at each position. This stacked image was used for object detection and photometry. 
The stacked images in each filter were aligned with the task {\it wregister} from IRAF, using the $i'$ band image as reference. 
We also combined the $g'$, $r'$ and $i'$ stacked images of each cluster to have a final $g'+r'+i'$ coadded image, using  IRAF's task {\it imcombine} (each image was scaled by the mean before being added).
This coadded image was used to visually inspect for gravitational arc candidates.

\begin{figure}
\centering
\includegraphics[scale=0.42]{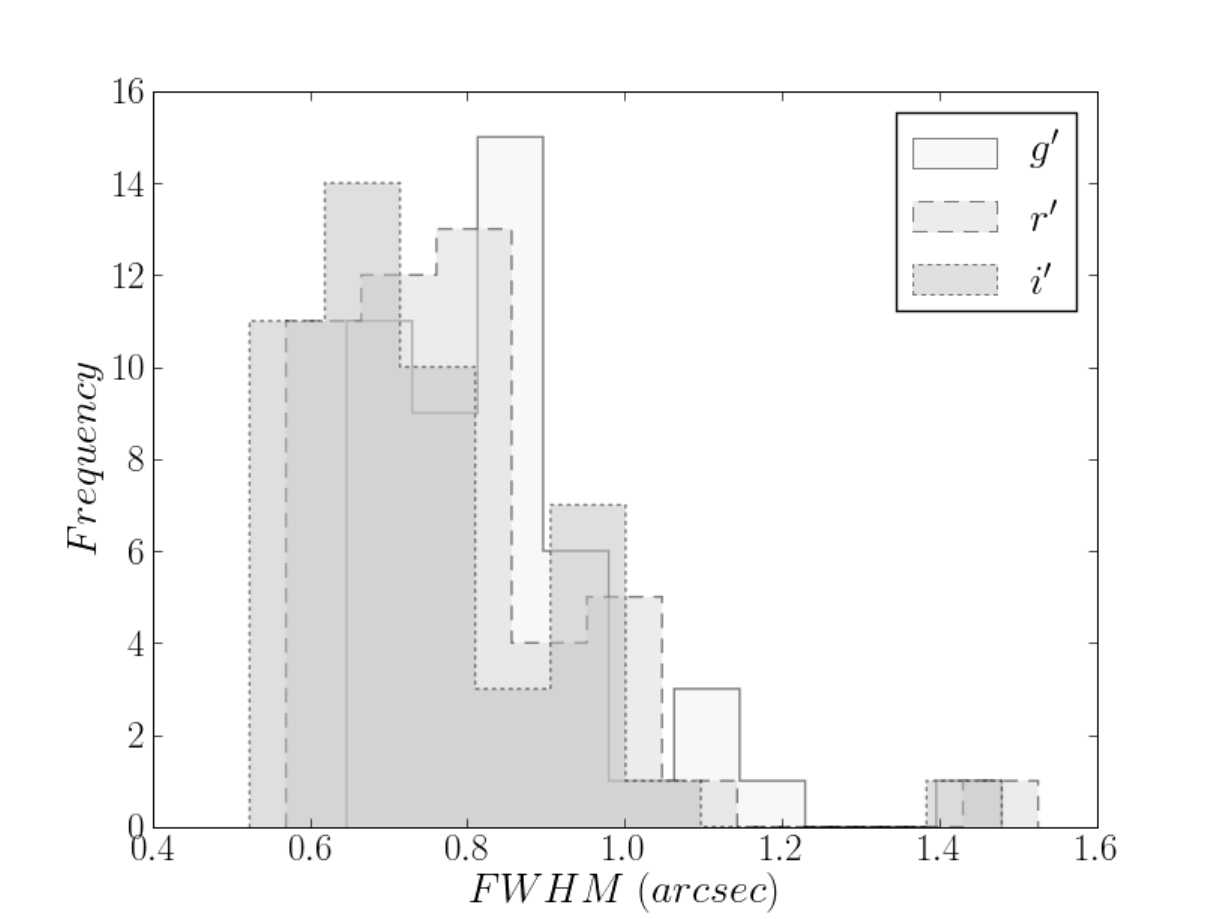}
\caption{Seeing FWHM distribution of the SOGRAS fields.}
\label{seeing_histo}
\end{figure}

We measured the seeing in the stacked images for each band using the {\it imexamine} IRAF task. 
The distribution is shown in Fig.~\ref{seeing_histo}. Clearly most of the images do satisfy our seeing constraint, at least in the $i'$ band, and are therefore well suited for finding arcs. Only two clusters have seeing larger than $1''$ in that band. They were observed in windy conditions and their images have relatively poor seeing. The median seeing for all images is $0.83''$, $0.76''$ and $0.71''$ in the $g'$, $r'$ and $i'$ bands respectively.

\subsection{Astrometric calibration}

For the construction of the world coordinate system (WCS) of each stacked image, we used The Guide Star Catalog, Version 2.3.2, GSC2.3-STSCI \citep{gsc} and pattern-matched the positions of the stars against those in the SOGRAS fields, producing a list with the RA, Dec and cartesian ($x$ and $y$) coordinates of the selected stars. For this process we selected a subsample of the brightest (but not heavily saturated) stars yielding typically about 20 stars for each SOGRAS field. The residuals in the astrometric solutions for RA and Dec are of the order of $0.15''$, which is sufficient to enable a proper positional matching to SDSS for the photometric calibration.

\subsection{Photometric calibration}\label{photocalib}

We used public SDSS data to calibrate the photometry of our stacked images. 
Following the standard SDSS system, we hereafter refer to our calibrated data as $g$, $r$ and $i$ magnitudes. Most fields 
we observed belong to the SDSS Stripe 82 and therefore have easily available and very accurate photometric calibrations \citep{Ivezic07}. 

The technique we employed to calibrate our sample was as follows.  
The first step was to select bright but unsaturated stars from SDSS DR7 \citep{Ivezic07} in the SOGRAS fields. For each field, 
we detected objects with SExtractor Version 2.8.6 \citep{Bertin96} 
and defined as stars objects 
whose star-galaxy classification parameter CLASS\_STAR, as measured by SExtractor in the $i$ band image, was larger than 0.85.
Those were matched to the stars in  
the SDSS catalog, yielding typically about 10 stars per SOGRAS field. 
Then we computed the mean offset between the magnitudes measured with SExtractor's 
automatic aperture (MAG\_AUTO) and their corresponding SDSS magnitudes (MODEL\_MAG), and used this offset as the zero-point 
of the magnitude scale of each filter for that field. 
The mean (dispersion) values are $31.46~ (0.15)$, $31.35 ~(0.10)$ and $30.90 ~(0.10)$, for $g$, $r$ and $i$ filters, respectively.

Given the lack of actual photometric standards in our calibration process
and the fact that our focus is on extended objects, we refrained from
adopting colour terms in the calibration, whose amplitude is likely smaller
than the uncertainties in the galaxy photometry. 
\subsection{Galaxy photometry}
\label{gal_phot}

We used SExtractor to identify sources and measure their 
magnitudes. Based on provided detection thresholds, SExtractor determines 
the background around each source
and whether a given pixel belongs to the source or to the background. SExtractor automatic aperture photometry (MAG\_AUTO) was adopted in this work. It is based on flexible elliptical apertures around every detected object and is 
intended to give a  
precise estimate of total magnitudes, at least for galaxies. 

An automated pipeline in python was created to expedite the process of object finding, photometry (including application of photometric calibration described earlier) and catalogue construction. This pipeline uses functions from \texttt{SLtools}\footnote{The \texttt{SLtools} library is available at http://che.cbpf.br/sltools/}, a library for image processing, catalogue manipulation and strong lensing applications (Brandt et al., in preparation). 
 In the following we provide a brief description of this pipeline.

For each SOGRAS field, our pipeline runs SExtractor separately for the $g$, $r$ and $i$ stacked images, using the previously determined zero-points. Sources with more than 10 contiguous pixels and whose flux exceeds $2 \sigma$ above the sky background were considered as real detections\footnote{The remaining parameters used for object detection and catalogue generation with SExtractor can be obtained from the configuration file, which is available upon request to the corresponding author.}. Therefore, the resulting AUTO magnitudes for the same galaxy in different 
filters are meant to quantify their total fluxes.  We did experience with flux 
measurements in the same aperture and area, using SExtractor in the dual image mode, taking the $i$ band image as reference image (i.e., using the aperture defined
in $i$ band image for measuring magnitudes on the images taken in $g$ and $r$), but this procedure
resulted in systematics in colours $10-50\%$ larger than those here presented (see the next section).

In section \ref{catalogs} we describe the resulting catalogs in more detail. 

\subsection{Photometric Quality Assessment}
\label{QA}

In order to assess the quality of the SOGRAS photometry, we again used the well-calibrated data from SDSS, given the overlap of the SOGRAS fields with the SDSS footprint. We began by performing an object matching for each SOGRAS field, as was done in the selection of stars for photometric calibration. 

\begin{figure*}
\begin{minipage}[b]{1.0\linewidth}
\centering
\includegraphics[scale=0.42]{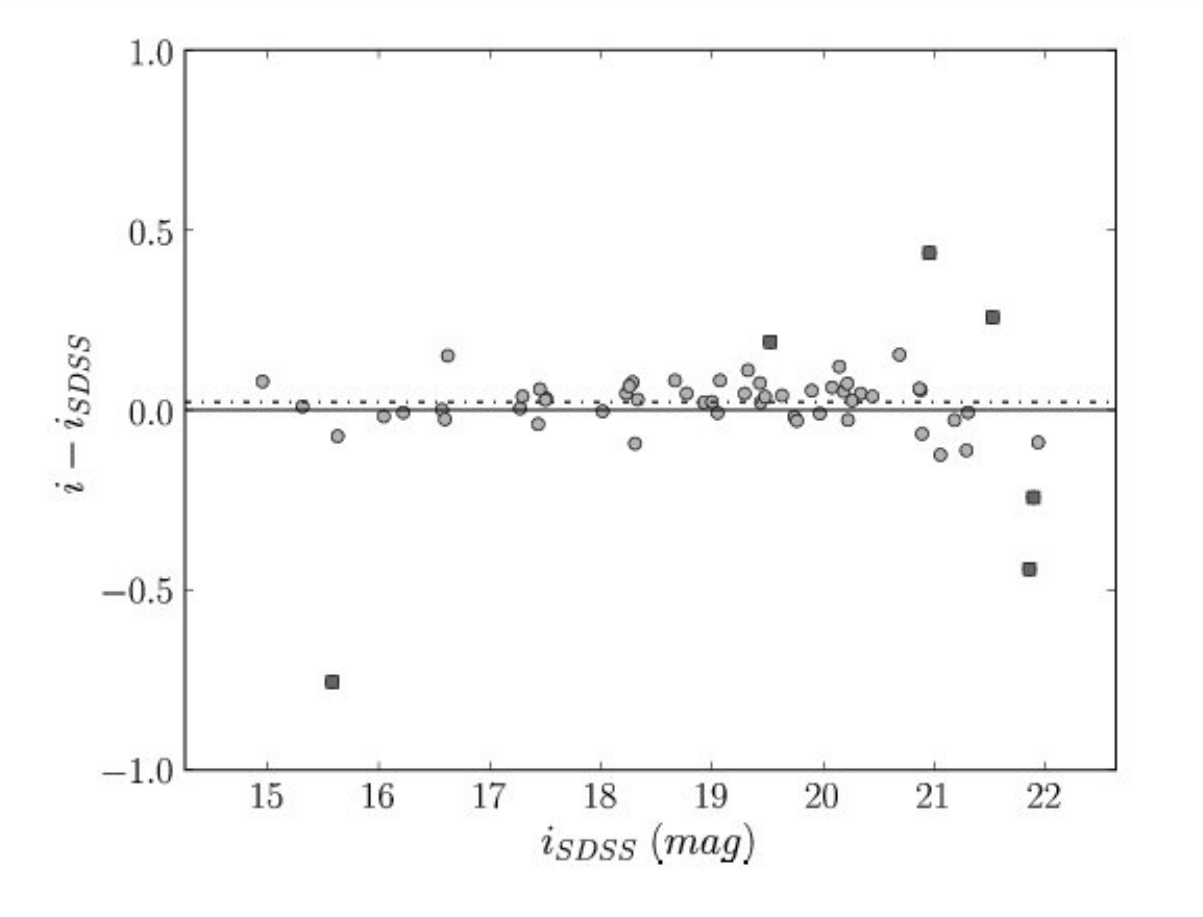}
\includegraphics[scale=0.42]{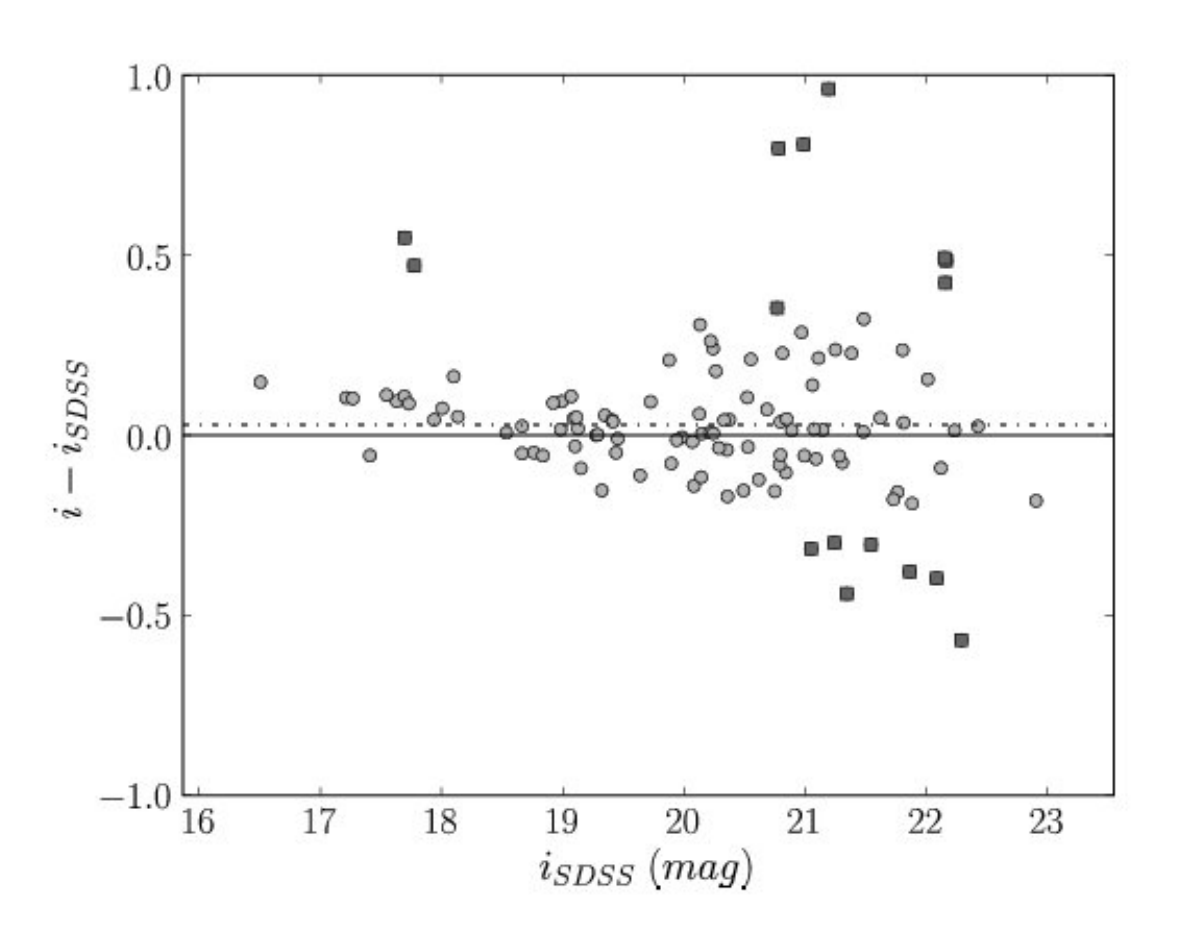}
\includegraphics[scale=0.42]{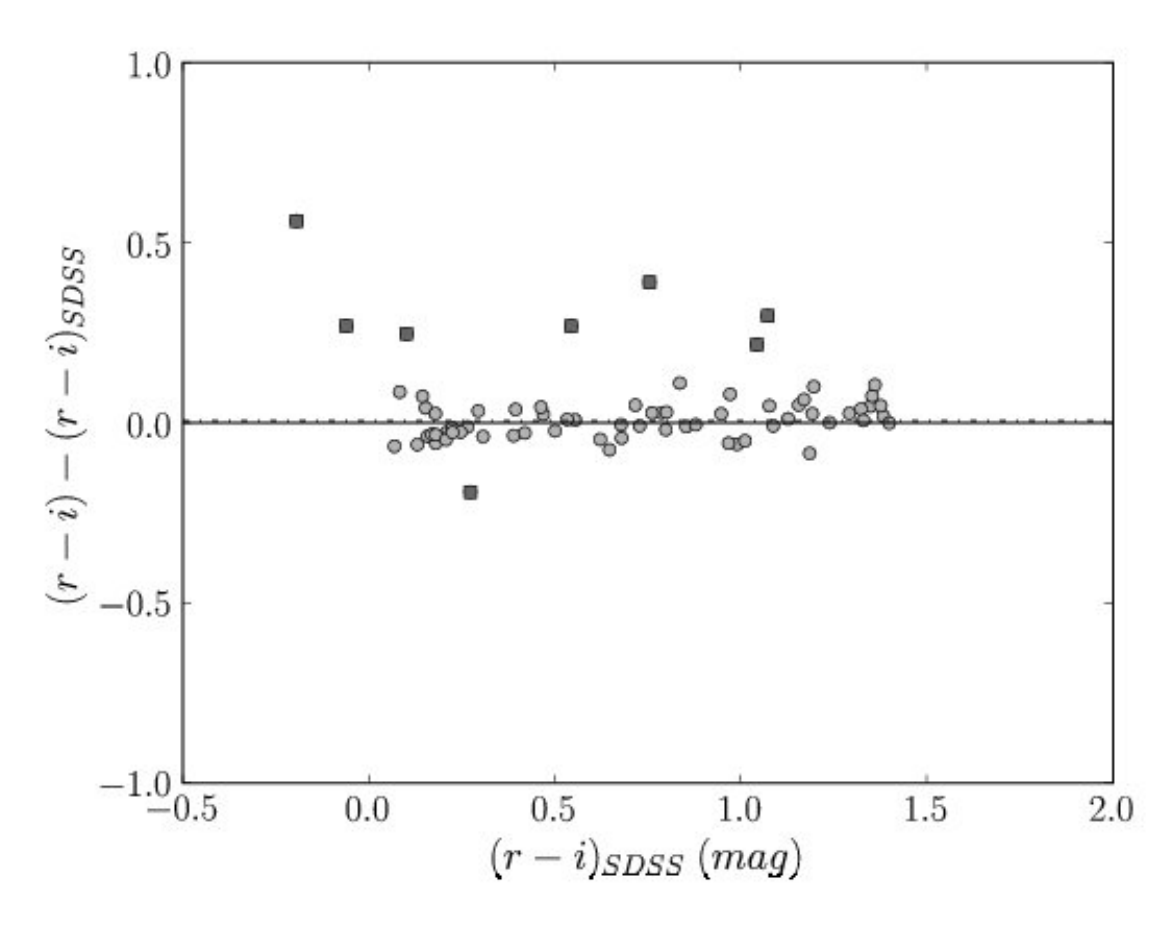}
\includegraphics[scale=0.42]{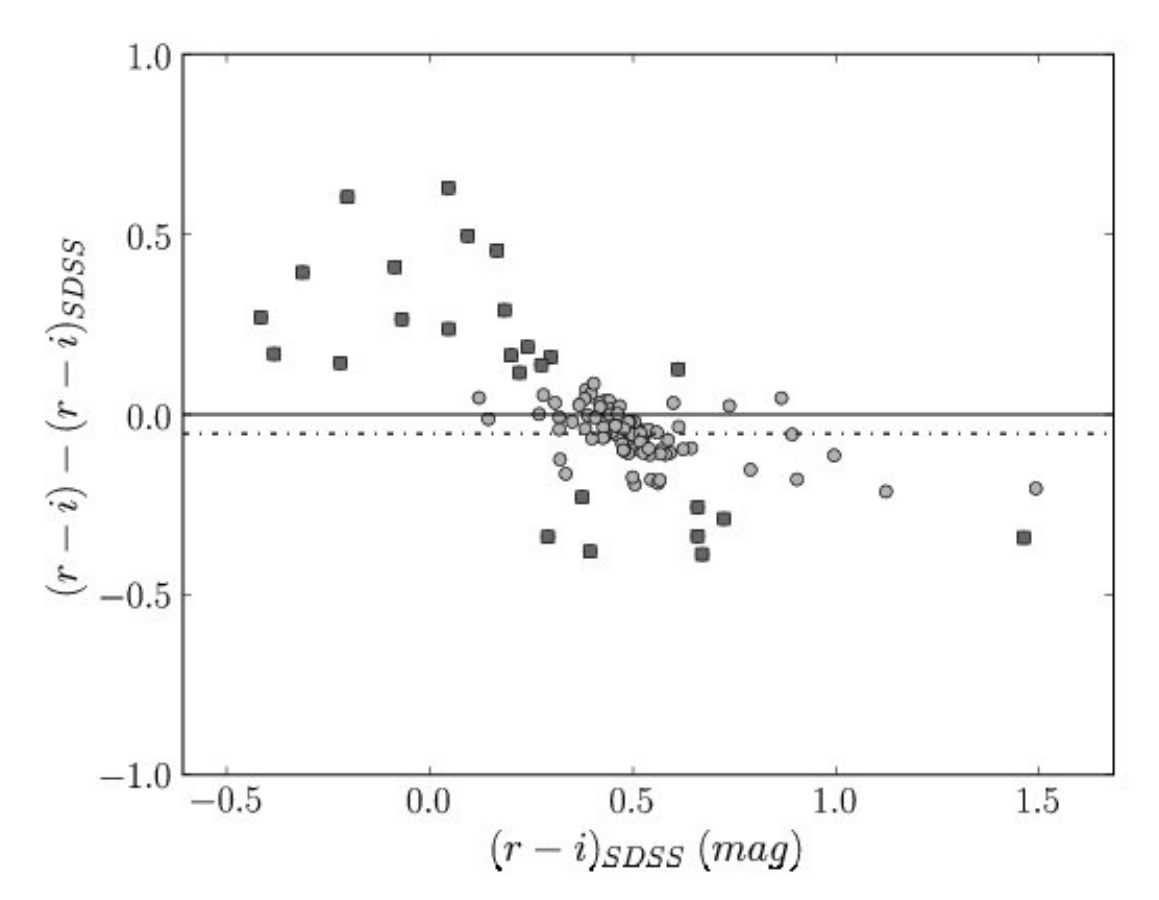}
\caption{Differences in $i$ band magnitude (upper panels) and $(r-i)$ colour (lower panels) between SOGRAS and SDSS for both stars (left panels) and galaxies (right panels) in the field around the cluster SOGRAS0850+0015. Only objects with $r < 22.5$ were used in the plots. The solid line corresponds to zero residual. The square points were $2.5 \sigma$ clipped out before computing the mean residual (dashed line).} 
\label{qa_sa303}
\end{minipage}
\end{figure*}

We compare our magnitudes and colours to those from SDSS in order to assess our photometric errors and their dependence on $S/N$ level. In Fig. \ref{qa_sa303} we show, as an example, the comparison of $i$ band magnitudes and $(r-i)$ colours for both stars (shown on the left panels) and galaxies (right panels) in the field of the cluster SOGRAS0850+0015. At the bright end, the scatter in the plots is dominated by the differences in the way magnitudes were measured in SOGRAS and SDSS, and by
the residuals in photometric calibration. On the other hand, at the faint end of the plots, the larger scatter is probably caused by the low $S/N$ levels of these objects, specially in SDSS, which is shallower than SOGRAS. 
No significant systematics is seen in the stellar
photometry, indicating that the photometric calibration is effective. After
applying a 2.5-$\sigma$ clipping to eliminate outliers, we find a mean
offset of $\left<i-i_{SDSS}\right> = 0.02$ for 
$i < 19$. The {\it rms} SOGRAS-SDSS residual is $i=0.04$ in the same range. Stellar colours have slightly
larger systematic residuals $(r-i)-(r-i)_{SDSS} = 0.05$ with an {\it rms} value of
0.04 when bright stars are considered. These values are typical of the other 
fields.

\begin{figure*}
\begin{minipage}[b]{1.0\linewidth}
\centering
\includegraphics[scale=0.4]{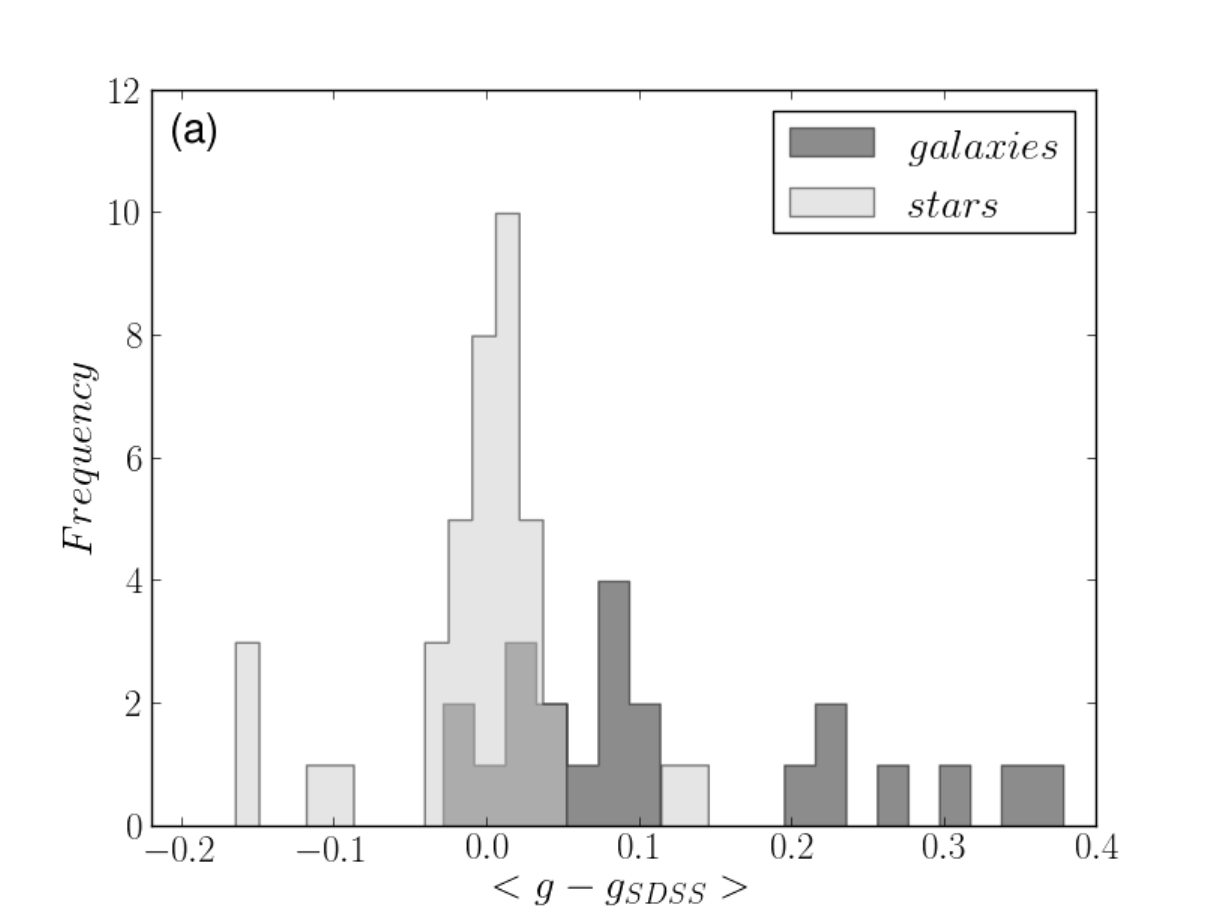}
\includegraphics[scale=0.4]{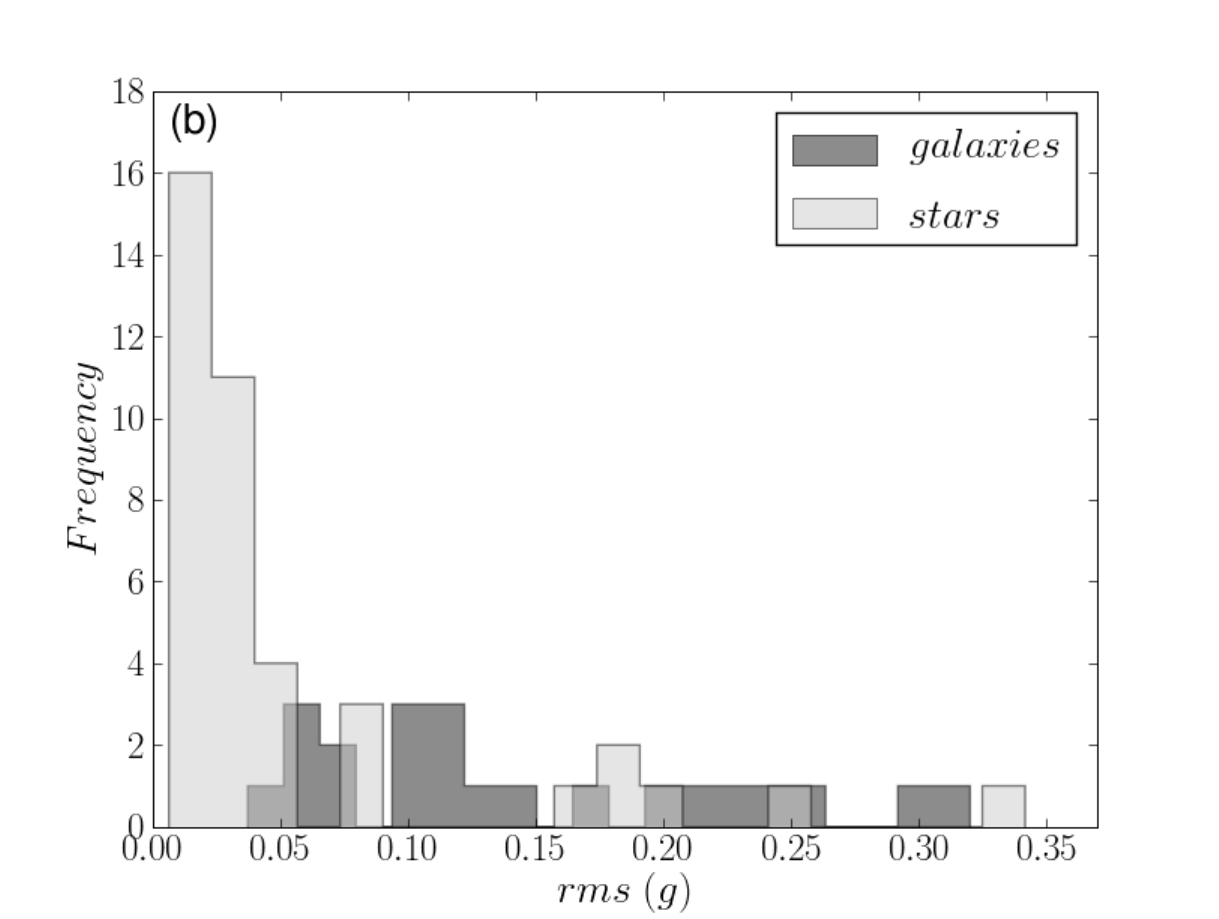}
\includegraphics[scale=0.4]{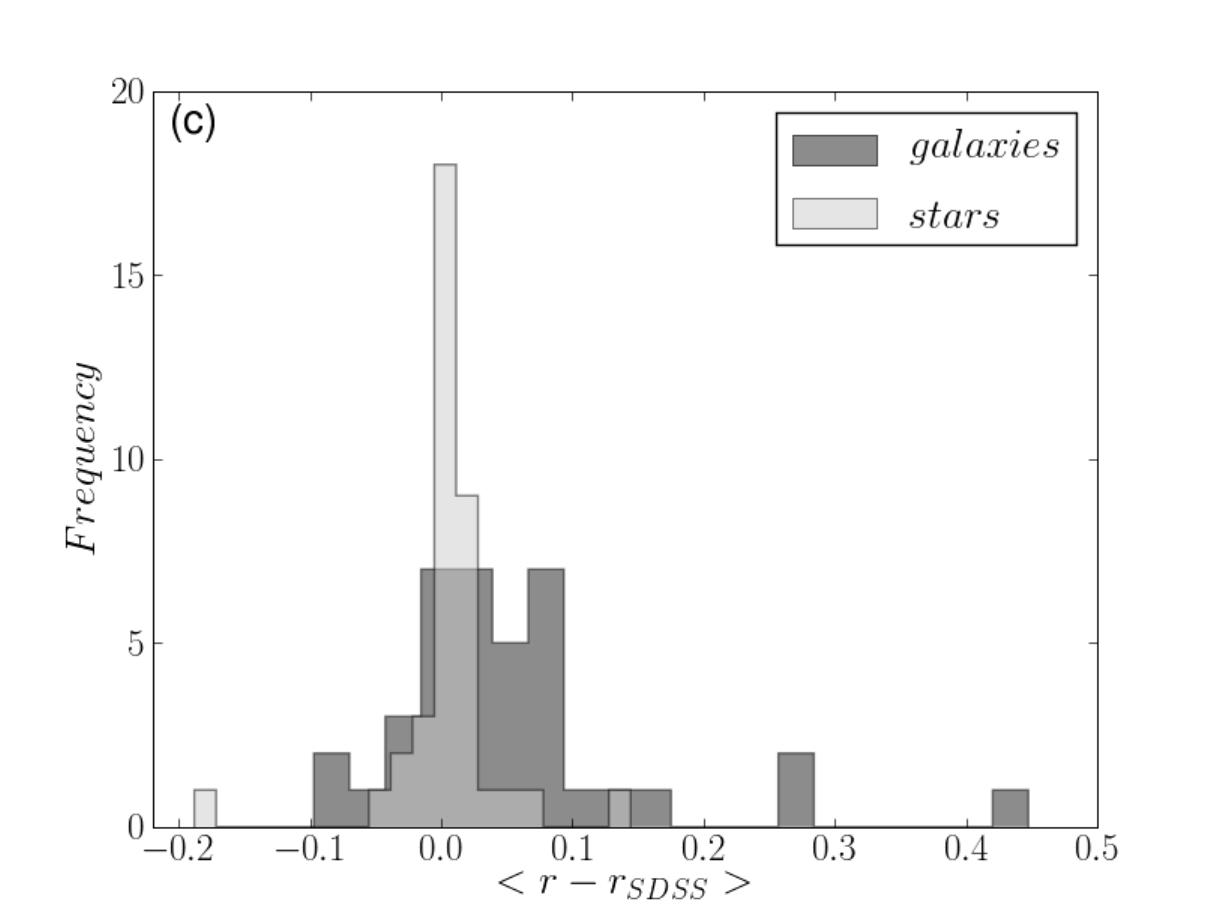}
\includegraphics[scale=0.4]{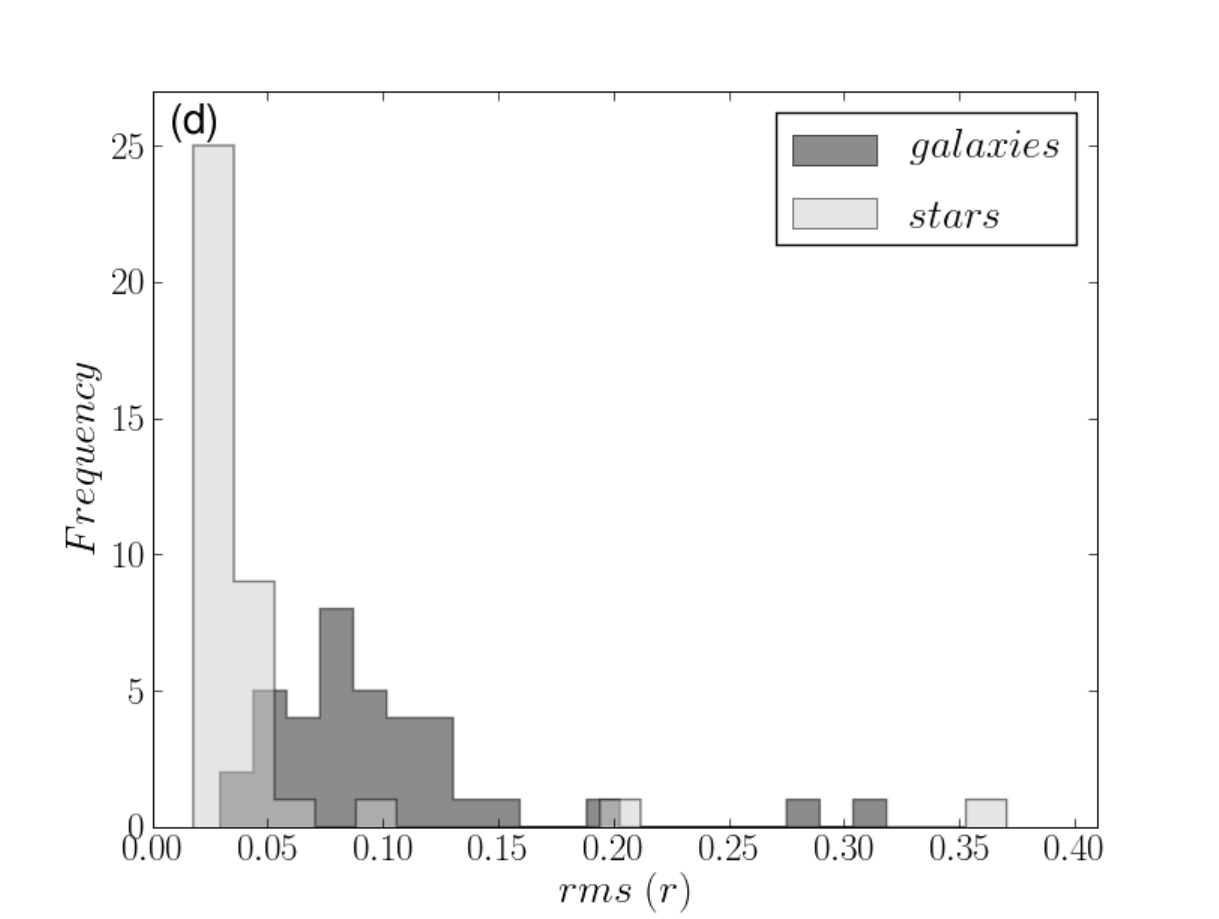}
\includegraphics[scale=0.4]{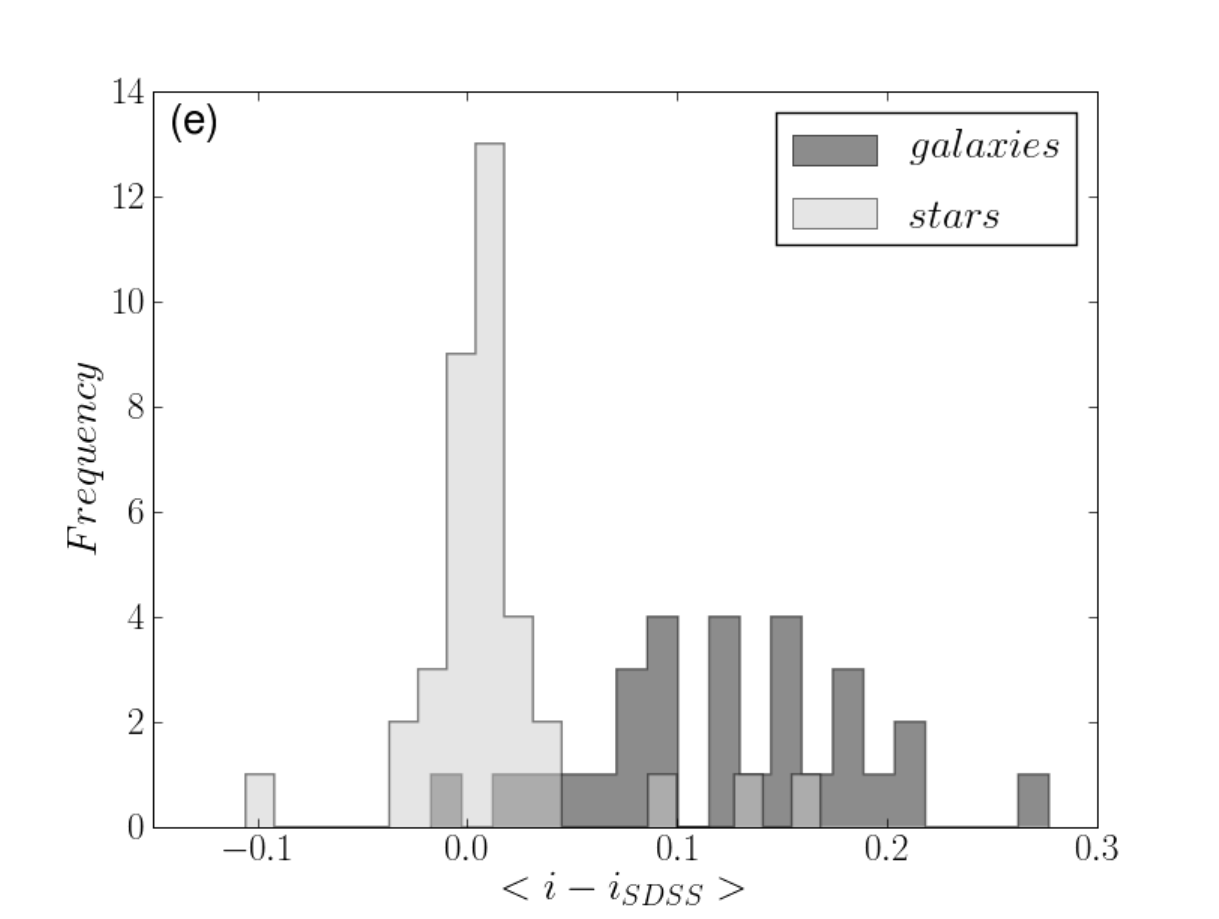}
\includegraphics[scale=0.4]{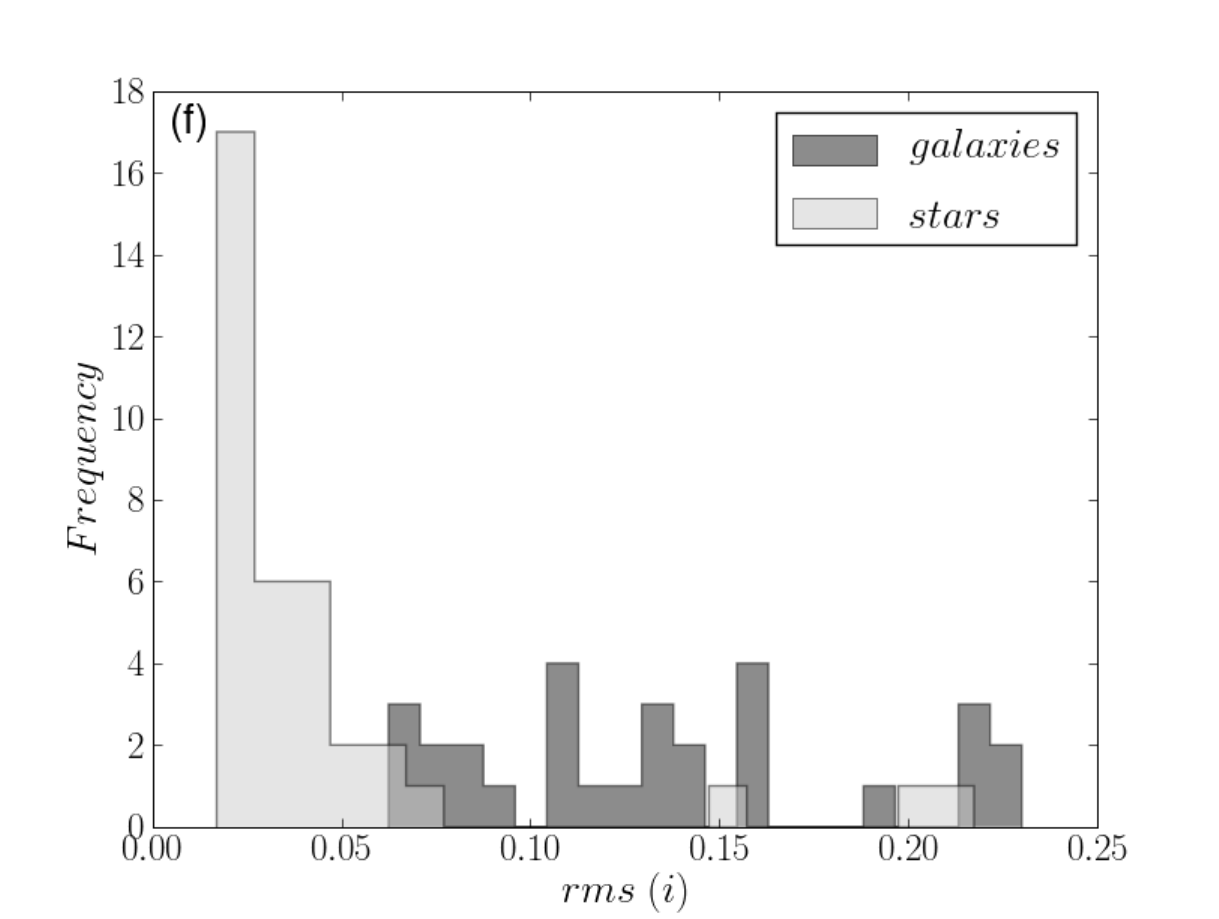}
\caption{Panel (a): distribution of mean $g$ band magnitude differences 
between our photometry and that of SDSS for bright stars (solid) and 
galaxies (dashed). Panel (b): distribution of {\it rms} $g$ band magnitude differences 
between our photometry and that of SDSS for bright stars (solid) and 
galaxies (dashed). See equation \ref{rms} and text for details. Panels (c) and (d): 
same as in (a) and (b) but now for the $r$ band. Panels (e) and (f): same as in (a) and (b) but now for the $i$ 
band. }
\label{calib_hist}
\end{minipage}
\end{figure*}

As for the galaxies, the mean offsets in $i$ magnitudes and $(r-i)$ colours 
depicted in Fig. \ref{qa_sa303} are comparable to those of stars. But the 
scatter is larger as can be attested from
a visual inspection of Fig. \ref{qa_sa303}. This reflects the
difficulty in measuring their total fluxes, specially for faint
objects, and also the differences in the SOGRAS and SDSS PSFs.  The {\it rms} residuals are 
0.06 both for magnitudes and for colours.
The scatter plot in colours also reveals a systematic trend of SOGRAS galaxies
being redder (bluer) for blue (red) colours, which is another way of concluding
that the SOGRAS colour distributions are narrower than those from SDSS. This
is just as expected from the higher photometric precision and image quality from
our SOAR data.

We now extend our photometric quality assessment to the other SOGRAS fields.
In Figure \ref{calib_hist},  the panels on the
left show the distribution, over all fields, of the mean difference between
our magnitudes and those from SDSS using bright sources only. The
panels on the right show the {\it rms} difference, computed as follows:
\begin{equation}
rms = \frac{1}{\sqrt{2}} \left(\frac{\sum_{j=1}^{N} (m_j-m_{j,SDSS})^2 }{N}\right)^{1/2},
\label{rms}
\end{equation}
where $m_j$ are the individual magnitude measurements from either surveys in a
given field and $N$ is the number of bright sources in that field. 
By {\it bright sources} in this figure we mean those with $g < 20$, $r < 20$, 
$i < 19$. The $\sqrt{2}$ factor in the expression for the $rms$ 
accounts for the fact that errors in {\it both} 
SDSS and SOGRAS
contribute in quadrature to this statistic. Therefore, we are here taking the
relatively conservative approach of that both photometric samples are of
equal precision, which may lead to an overestimate of the SOGRAS random
photometric errors. 
The solid (dashed) histograms in each panel are for the stars (galaxies). 
Each row corresponds to a given filter. Only fields with at least 5 bright
sources were included in the histograms.

The mean photometric offsets for the stars 
are well centered around $\left<m-m_{SDSS}\right> = 0$, with very few SOGRAS fields
having $\vert \left<m-m_{SDSS}\right> \vert > 0.1$. The global median values of these offsets are $0.002$, $0.006$, 
and $0.005$ for $g$, $r$ and $i$, respectively. This result shows that our calibration was successful and 
consistent for the whole survey. The mean offset for
galaxies is clearly larger and systematically positive. However, the typical
galaxy systematics is still constrained to $\left<m-m_{SDSS}\right> \le 0.2$ 
in most cases. This larger 
systematics reflects the complexity of measuring galaxy fluxes and is likely 
caused by the differences in measuring methods applied to SOGRAS (MAG\_AUTO) 
and SDSS (MODEL\_MAG). Notice that the mean galaxy offset also tends
to be larger in the $i$ band, likely as the result of fringing residuals 
accummulated on galaxy angular scales.

The $rms$ plots bear information on the random rather than systematic effects.
Since they are also restricted to bright objects, we can estimate the 
photometric uncertainty in our calibration using the stellar $rms$ of a typical field:
$rms(g) \simeq 0.03; rms(r) \simeq 0.03; rms(i) \simeq  0.03$. For the galaxies, these
typical values are larger, $rms(g) \simeq 0.11; rms(r) \simeq 0.09; rms(i) \simeq  0.13$,
since they incorporate the effect of different magnitude definitions
on top of that from calibration.

\begin{figure}
\centering
\includegraphics[scale=0.42]{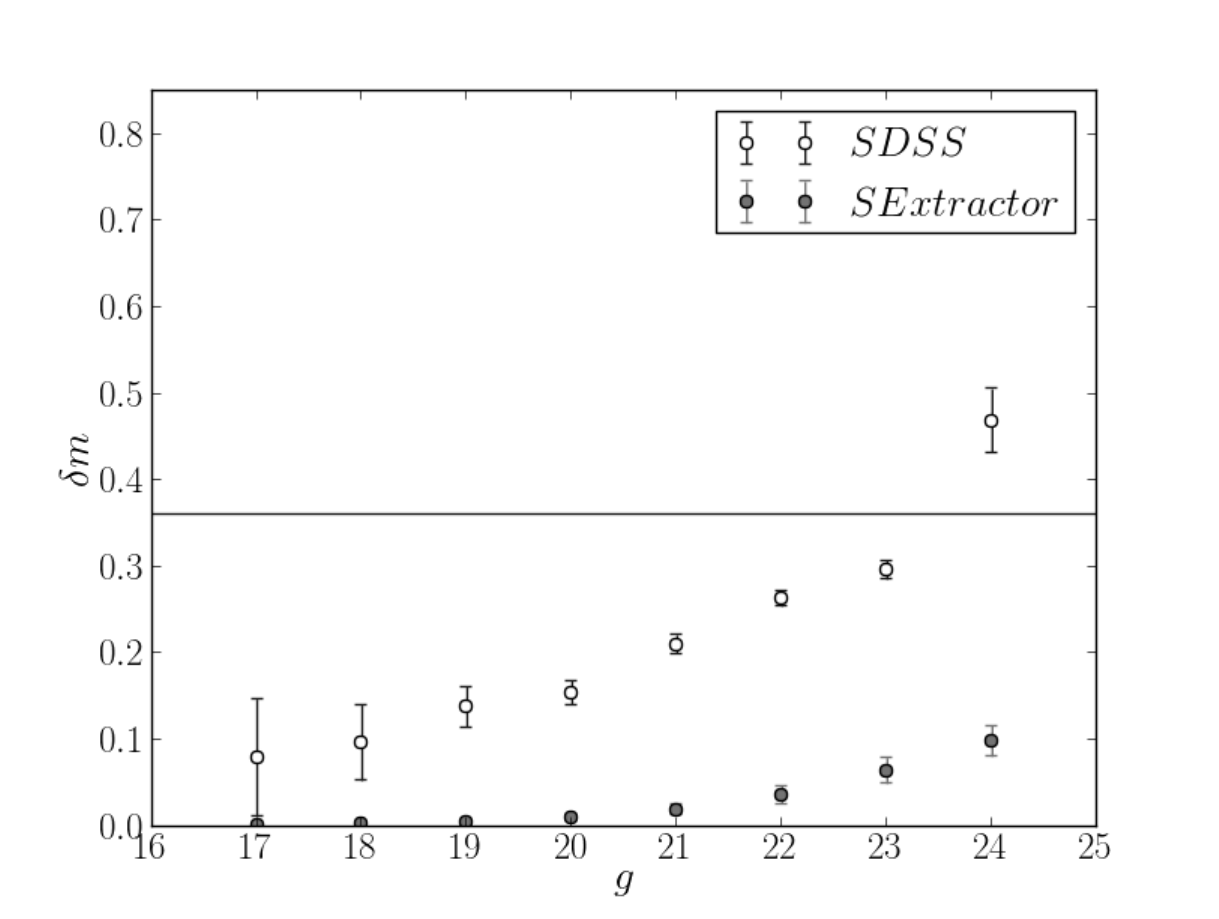}
\caption{Magnitude uncertainties in $g$ in each bin estimated as the $rms$ residuals relative to SDSS divided by $\sqrt{2}$ (open points) and estimated as the mean value of the errors provided by SExtractor (MAGERR\_AUTO, filled points). For the first case, the errorbars were estimated using a bootstrap resampling technique, while for the second case, they correspond to the dispersion around the mean. The $\delta_m = 0.36$ 
line indicates the $S/N=3$ level. }
\label{sigma_g}
\end{figure}

\begin{figure}
\centering
\includegraphics[scale=0.42]{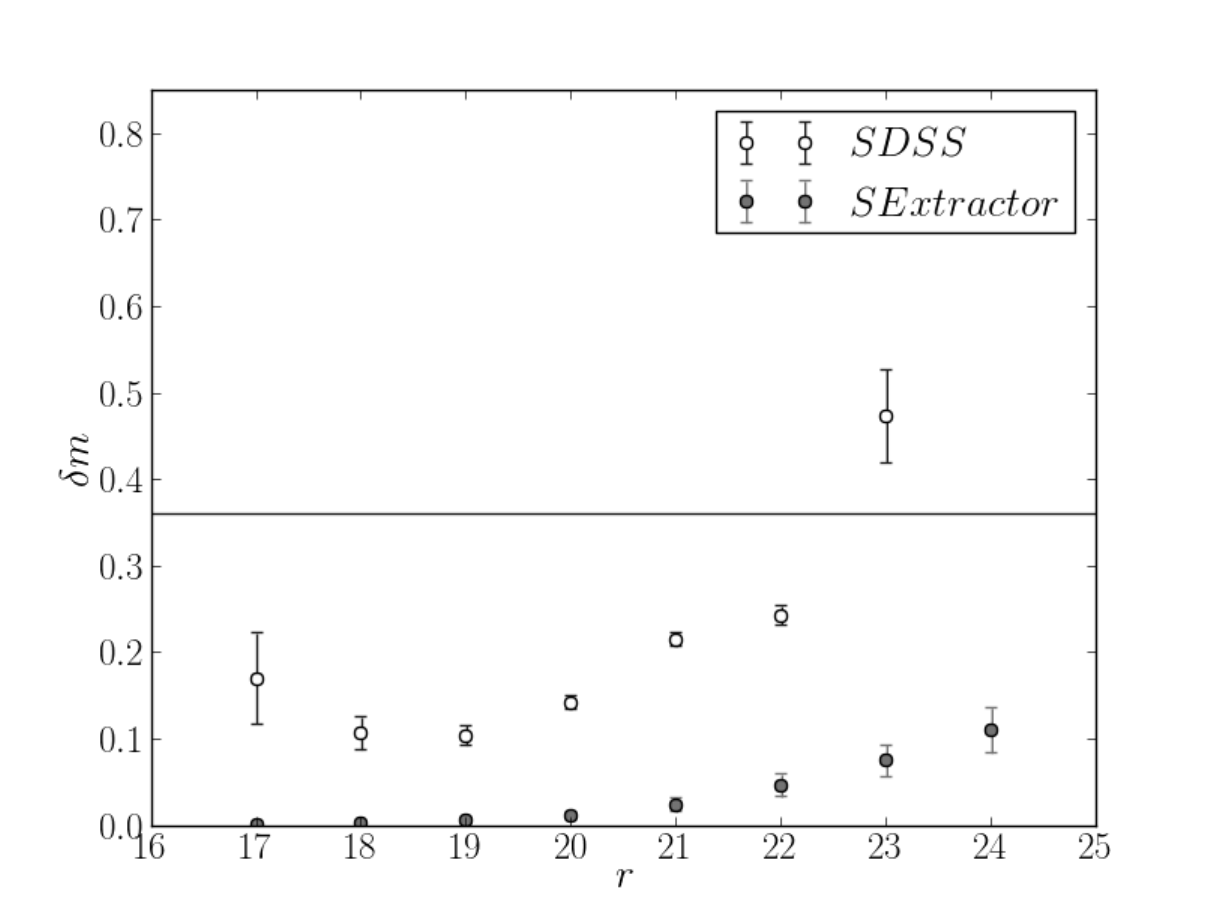}
\caption{Same as Figure \ref{sigma_g}, but for the $r$ band.}
\label{sigma_r}
\end{figure}

\begin{figure}
\centering
\includegraphics[scale=0.42]{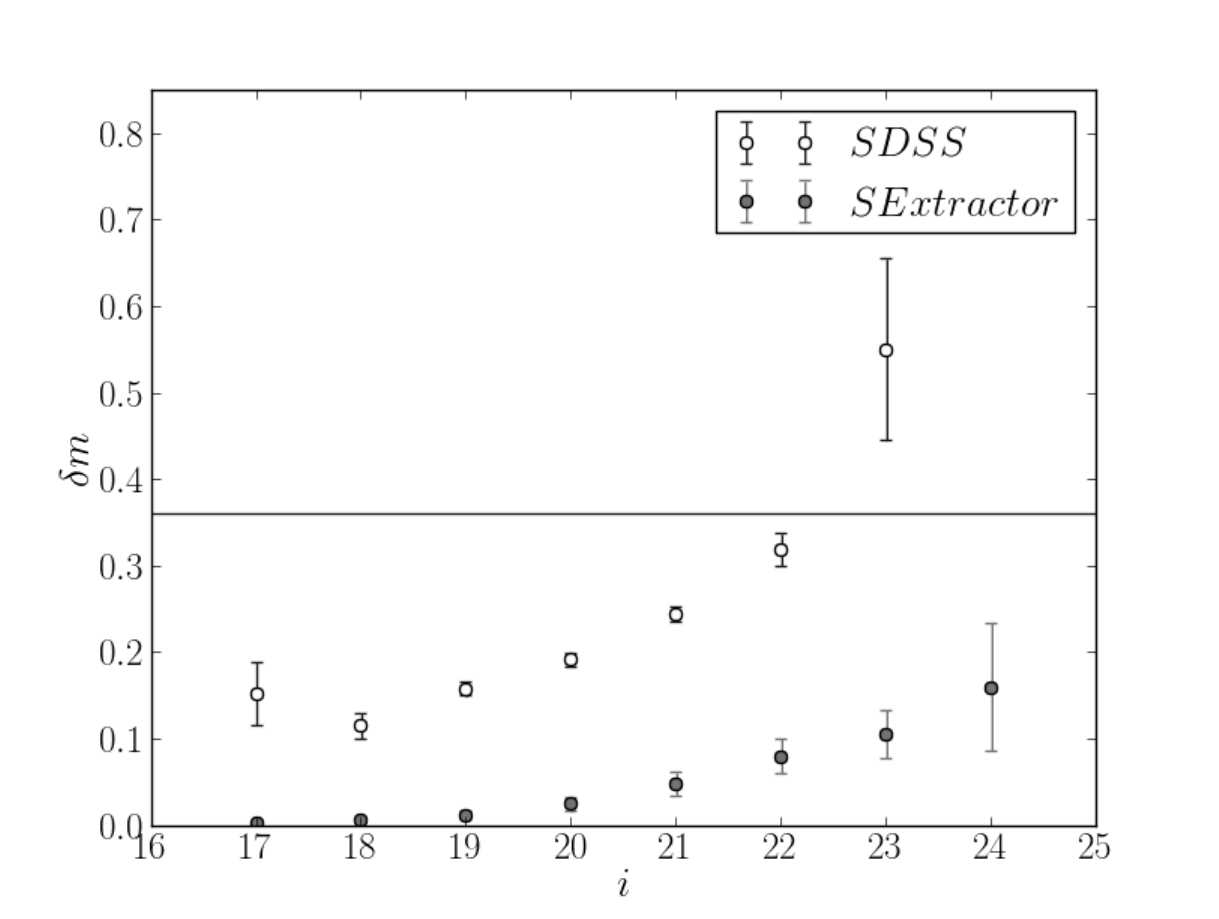}
\caption{Same as Figure \ref{sigma_g}, but for the $i$ band.}
\label{sigma_i}
\end{figure}

In Figures \ref{sigma_g}--
\ref{sigma_i}, we assess
the SOGRAS random photometric errors, $\delta_m$, as a function of $S/N$ level. The values
of $\delta m$ (open points) are computed as the $rms$ difference given in equation \ref{rms}, but in this case, the sum is over all the galaxies in a given magnitude bin and the $rms$ have been corrected for the systematic residual between SOGRAS and SDSS galaxy magnitudes in each field, whose distribution of values are shown as the darker histograms on the left panels of Figure \ref{calib_hist}. The uncertainties in $\delta m$ are estimated via bootstrap resampling of all the galaxies in each magnitude bin. 1000 different realizations of the data were constructed using this method. We also plot the mean error in the MAG\_AUTO values provided by SExtractor (MAGERR\_AUTO) for the galaxies in each magnitude bin (filled points). The error bars in this case are the dispersion around the mean value.

The figures show a systematic increase in the photometric uncertainties
as fainter galaxies are considered. The magnitude uncertainties based on the SExtractor errors are much smaller than the ones based on $rms$ residuals relative to SDSS. This in part reflects our conservative assumption that both SOGRAS and SDSS are of equal accuracy. It may also reflect an underestimate of the photometric uncertainties by SExtractor. 
From these plots we infer that, in average, higher $S/N$ are achieved in the $g$, than in the $r$, and than in the $i$ bands. The detection limits for $S/N > 3$, which corresponds to $\delta_m > 0.36$, are found from the comparison with SDSS, being roughly $g \simeq 23.5$, $r \simeq 23$ and $i \simeq 22.5$.

\section{Galaxy and arc catalogs}
\label{catalogs}

\subsection{Galaxy catalogs}

We merged the catalogs resulting from the photometry (see section \ref{gal_phot}) of the 3 bands in a final object catalog, choosing the $i$ band catalogue as reference for object position. This means that we searched in the $g$ and $r$ band catalogs for the objects that are in the $i$ catalog. Non-matched objects received a flag $-99.99$ in the corresponding filter. The final catalogue was saved as a FITS table and contains the information on position, magnitudes, morphological parameters and star-galaxy classification of the detected objects. 
In Table \ref{catalog_labels}, we list the parameters in the final object catalog
and their corresponding definitions. 
This catalogue is publicly available upon request to the corresponding author.

\begin{table}
\caption{Column labels in the SOGRAS object catalog.}
\label{catalog_labels}
\begin{tabular}{l c} 
\hline
Parameter name & Definition \\
\hline
CLUSTER$\_$ID & Cluster ID\\

OBJECT$\_$ID  & Object number\\

RA$^a$  & Right ascension (J2000) \\

DEC$^a$  & Declination (J2000) \\

X$\_$IMAGE$^a$  & Object position along x \\

Y$\_$IMAGE$^a$  & Object position along y    \\

MAG$\_$AUTO$\_$G  & Automatic aperture magnitude in the $g$ band\\

MAGERR$\_$AUTO$\_$G  &  RMS error for AUTO magnitude in the $g$ band\\

MAG$\_$AUTO$\_$R  & Automatic aperture magnitude in the $r$ band\\

MAGERR$\_$AUTO$\_$R  & RMS error for AUTO magnitude in the $r$ band\\

MAG$\_$AUTO$\_$I  & Automatic aperture magnitude in the $i$ band\\

MAGERR$\_$AUTO$\_$I  & RMS error for AUTO magnitude in the $i$ band\\
  
THETA$\_$SKY$^b$  & Position angle (east of north) \\

ERRTHETA$\_$SKY$^b$  & RMS error for position angle \\

A$\_$IMAGE$^b$  & semi-major axis for second moments\\

ERRA$\_$IMAGE$^b$  &RMS error for semi-major axis\\

B$\_$IMAGE$^b$  & semi-minor axis for second moments \\

ERRB$\_$IMAGE$^b$  &RMS error for semi-minor axis \\

ELLIPTICITY$^b$   & Ellipticity (1 - B$\_$IMAGE/A$\_$IMAGE)\\

CLASS$\_$STAR$^a$  &  Star-galaxy classification \\

FLAGS$^a$   & Extraction flags  \\
\hline
\end{tabular}
$^a$Measured in $i$ band\\
$^b$Measured in $r$ band\\
\end{table}

\begin{figure*}
\centering
\begin{minipage}[b]{1.0\linewidth}
\includegraphics[scale=0.45]{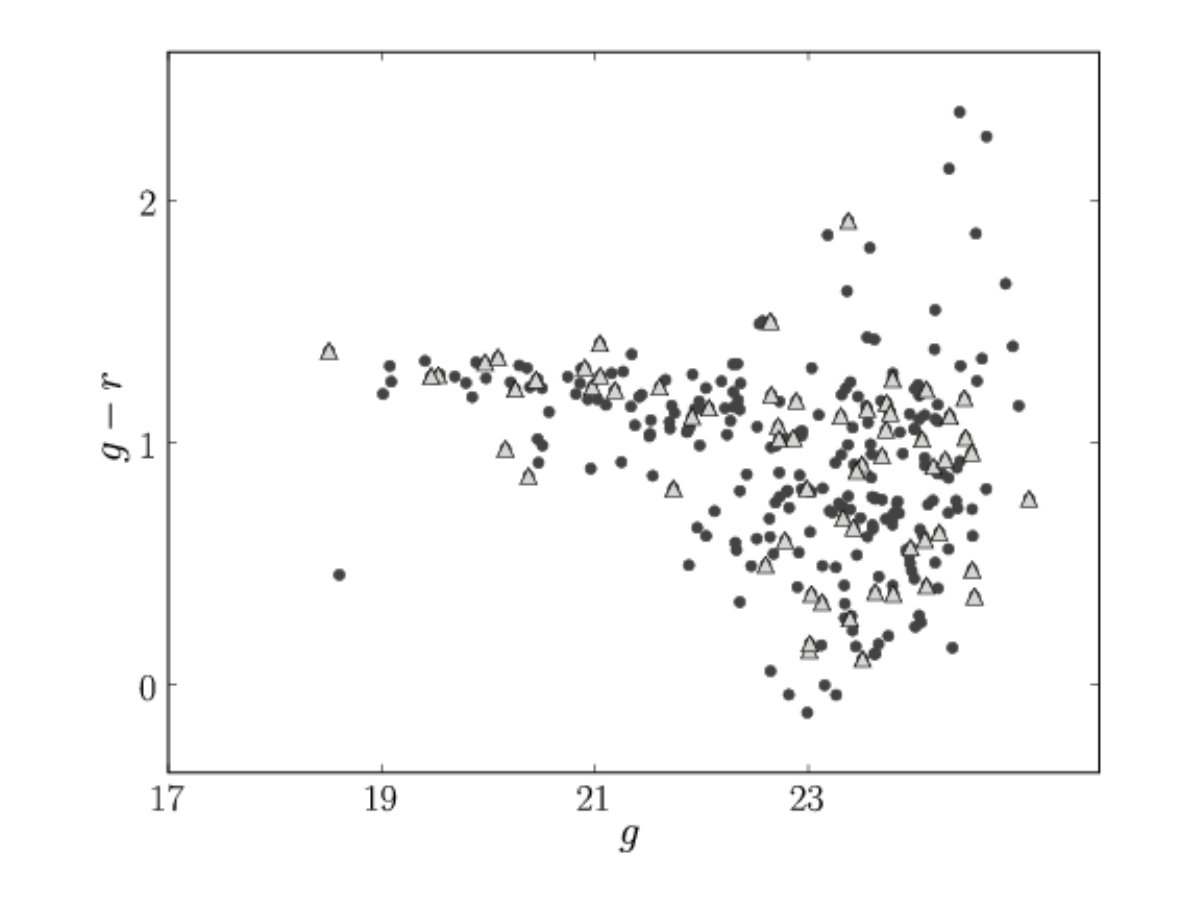}
\includegraphics[scale=0.45]{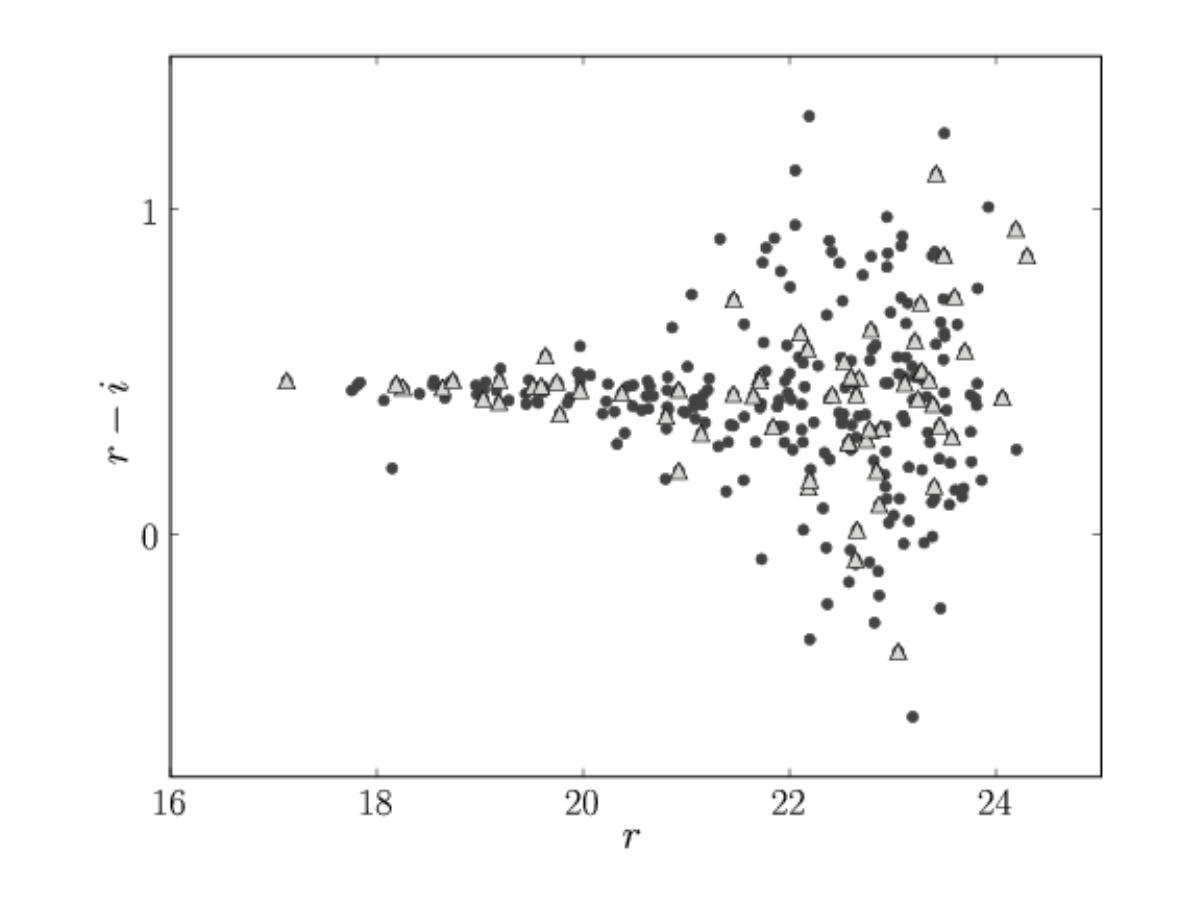}
\includegraphics[scale=0.45]{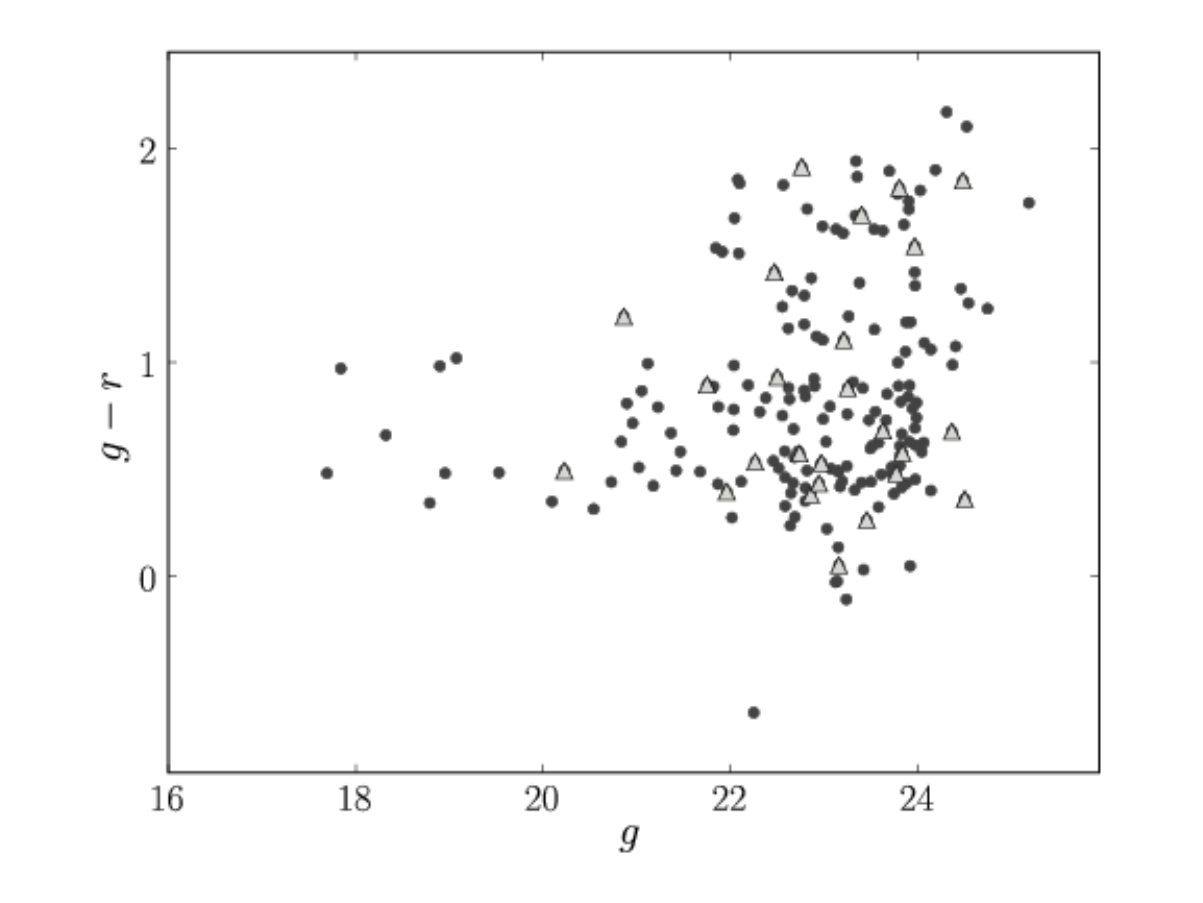}
\includegraphics[scale=0.45]{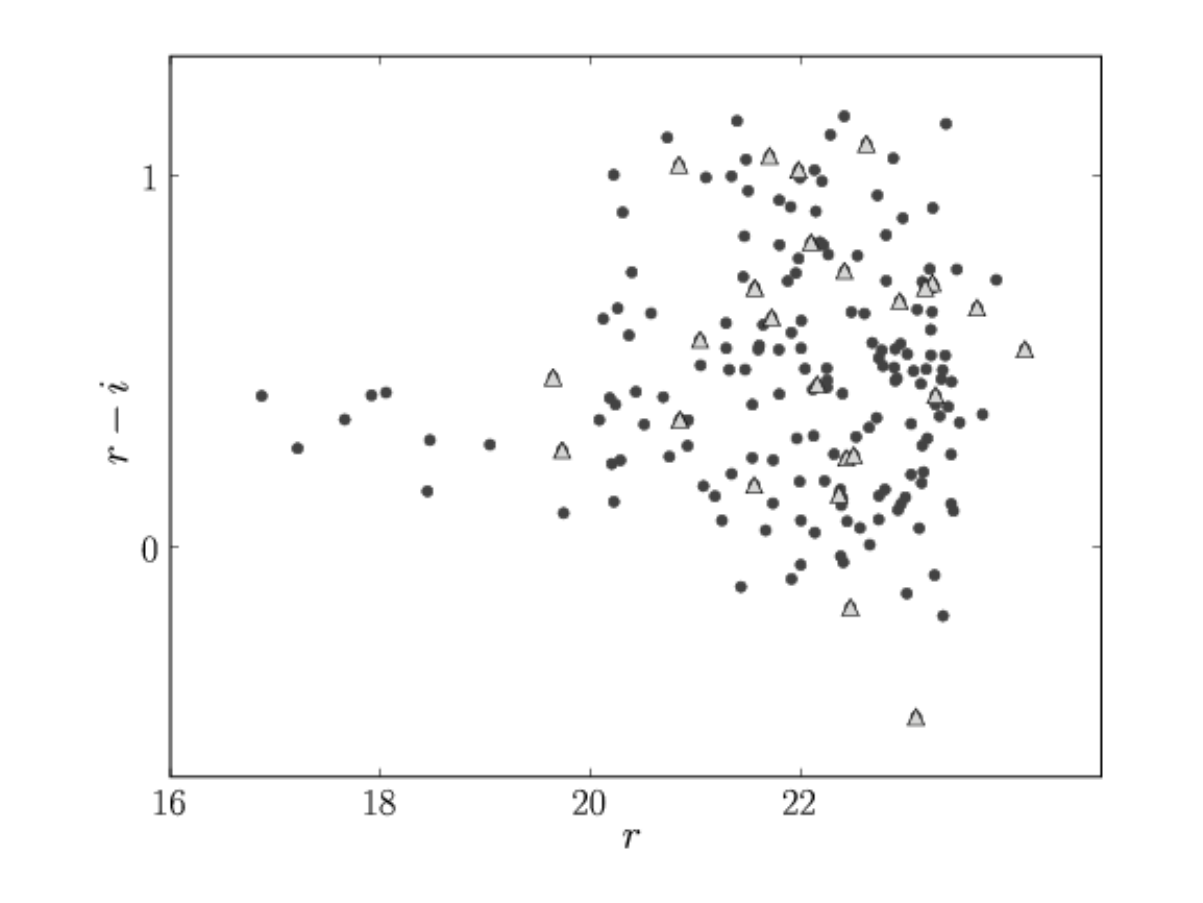}
\caption{Colour-magnitude diagrams of clusters SOGRAS0850+0015 (upper panels) and SOGRAS0202-0055 (lower panels). $(g-r)$ 
colours are shown on the left panels whereas
$(r-i)$ colours are on the right panels. The triangles indicate galaxies that are located in a circular region of $100''$  
(upper panels) and $60''$ (lower panels) around the cluster centre. Considering a flat $\Lambda$CDM model (with $\Omega_m=0.3, ~\Omega_{\Lambda}=0.7$ and $H_0=70$Km/s/Mpc) and the redshifts, these radii correspond to $0.33$Mpc and $0.4$Mpc, respectively.} 
\label{cmds}
\end{minipage}
\end{figure*}

In Figure \ref{cmds} we show colour-magnitude diagrams for two clusters,
one at the low-$z$ bin (SOGRAS0850+0015, $z=0.20$, upper panels) and the other with 
higher $z$ (SOGRAS0202-0055, $z=0.50$, lower panels). 
In order to reduce contamination from field galaxies in these diagrams, galaxies that are located in a circular region of $100''$  
(upper panels) and $60''$ (lower panels) around the cluster centres are shown with different symbols. Considering a flat $\Lambda$CDM model (with $\Omega_m=0.3, ~\Omega_{\Lambda}=0.7$ and $H_0=70$Km/s/Mpc) and the redshifts, these radii correspond to $0.33$Mpc and $0.4$Mpc, respectively.
The red sequence is more visible in the lower $z$ cluster, as expected.  
The colours of the red sequence also change slightly with $z$. The variation
seems to be systematically larger for $(g-r)$. For the relatively few high-$z$
clusters where the red sequence is clearly visible, it tends to have bluer
colours ($g-r \simeq 0.6$) than the low-$z$ ones, for which $(g-r) > 1$.
This is likely caused by the 4000\AA ~Balmer break affecting the $g$
filter at $z < 0.4$ but not the other passabands.

\begin{figure*}
\centering
\begin{minipage}[b]{1.0\linewidth}
\includegraphics[scale=0.45]{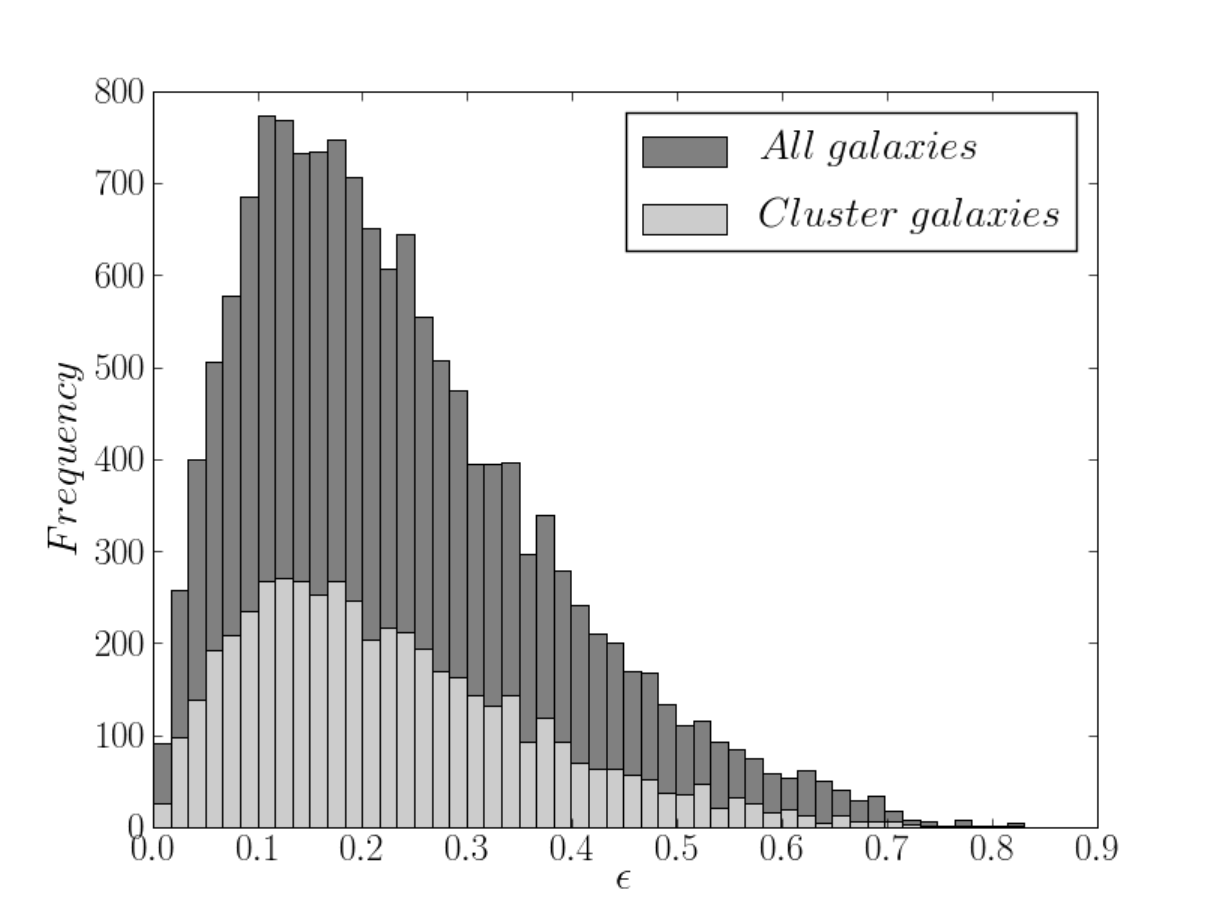}
\includegraphics[scale=0.45]{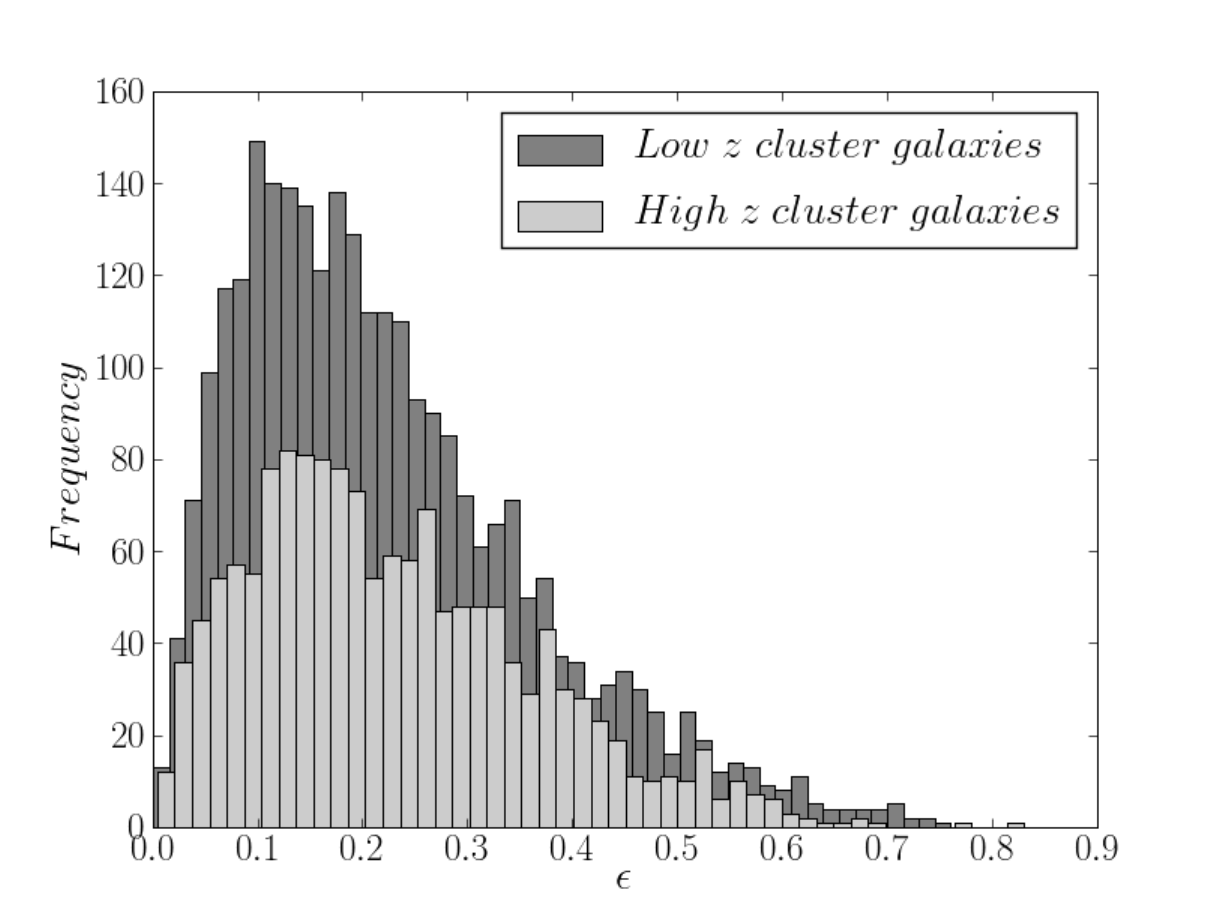}
\caption{Left panel: ellipticity distribution of all SOGRAS galaxies (all galaxies) and of SOGRAS galaxies which are located close to the clusters centre (cluster galaxies). Right panel: ellipticity distribution of SOGRAS galaxies which are located close to the clusters centre for low and high $z$ clusters. Galaxies were selected as objects with star-galaxy parameter CLASS\_STAR $<  0.85$, as measured in $r$ band images. }
\label{ellip_dist}
\end{minipage}
\end{figure*}

We chose to measure the morphology --- semi-major (A) and semi-minor (B) axes and position angles derived from weighted second moments in SExtractor --- in the $r$ stacked images to optimize the combination of $S/N$ and seeing. The seeing is not significantly degraded from the $i$ to the $r$ band (see Fig. \ref{seeing_histo}), while the $S/N$ increases from most objects (c.f. section \ref{QA}). Therefore, the $r$ band provides a balance between the higher $S/N$ of the $g$ band and the better seeing of the $i$ band. 
Figure \ref{ellip_dist} shows the distribution of the ellipticity $\epsilon= 1-B/A$ 
of the SOGRAS galaxies. The left panel shows the distribution of $\epsilon$ for all
objects classified as galaxies and for all of them which are located close to the centre
of the SOGRAS clusters (which we refer to as ``cluster galaxies'').  The latter objects were selected within circular regions around the cluster centres, whose angular radii were estimated visually. As in the case of figure \ref{cmds}, these cuts in angular separation were used just to reduce contamination from field galaxies.

Both distributions are peaked at $\epsilon \simeq 0.15$,
and are strongly skewed towards higher values. The peak position is affected by the relatively large errors in $\epsilon$ for round objects, given the constraint that $\epsilon \ge 0$ always. Typical errors for round objects ($\epsilon \le 0.15$) are of the order of $\sigma_{\epsilon}\simeq 0.07$. For more eccentric objects ($\epsilon > 0.15$), the errors are typical of $\sigma_{\epsilon}\simeq 0.04$. This behaviour is qualitatively
confirmed by analyses of early-type galaxies in nearby clusters 
\citep[e.g.,][]{Fasano10}. In the right panel we present the distribution of $\epsilon$ for SOGRAS galaxies that  are located close 
to the clusters centre for low and high $z$ clusters. The similarity between the distributions for low and high $z$ clusters 
indicates that our shape determinations for high-$z$ objects are not strongly degraded by atmospheric seeing and that shapes were properly measured for the
galaxies in our catalog. 

\subsection{New arc system candidates}
\label{
}

We inspected all SOGRAS images in order to look for strong gravitational lens systems. We found 6 clusters (SOGRAS0321+0026, SOGRAS0328+0044, SOGRAS0014-0057, SOGRAS0041-0043,%
\footnote{This arc system was subsequently found by independent arc searches in CS82 data, both from a visual inspection of cluster images as well from an automated arc search on the CS82 footprint (More et al., in preparation).}  SOGRAS0940+0744 and SOGRAS1023+0413) that show clear evidence of arcs and 2 clusters (SOGRAS0219+0022, SOGRAS0202-0055) that show probable arcs. 
  
We identified 16 arc candidates close to the brightest members of 8 cluster cores and most of them show bluer colours than the central cluster galaxies. Four of them are giant arcs and have length-to-width ratio ($L/W$) larger than 7. The remaining candidates are arclets, i.e. have smaller $L/W$. The length and width of the arc candidates were visually estimated, using the ds9 software\footnote{\texttt http://hea-www.harvard.edu/RD/ds9/}. The length was obtained by summing the two segments that connect the extreme points to the arc geometric centre. The width corresponds to the distance between the arc ``borders'' along the perpendicular bisector of the segment connecting the arc extrema passing through the arc centre.

The position, magnitude, length and width of the arc candidates in the 6 most probable strong lenses are displayed in Table \ref{tab_arcs}. Contrary to the measurements of magnitude of other objects in the catalog, which were obtained with SExtractor, the magnitudes of the arc candidates were measured using the task {\it polyphot} from IRAF. This task computes the magnitudes inside polygonal apertures, providing more precise measurements of arc magnitudes, since it takes into account the arc shape. The polygons were visually defined and meant to incorporate the total flux of the arcs in each filter. 

In Figure \ref{arc_candidates} we show the candidate strong gravitational lens systems identified in SOGRAS data. Notice the systematically bluer colours of the arcs in comparison to that
of a typical central cluster galaxy in our range of redshifts, $(g-r) \simeq 0.8-1.0$ (see Table \ref{tab_arcs} and online colour version of Figure \ref{arc_candidates}).

\begin{figure*}
\centering
\begin{minipage}[b]{0.98\linewidth}
\includegraphics[scale=0.36]{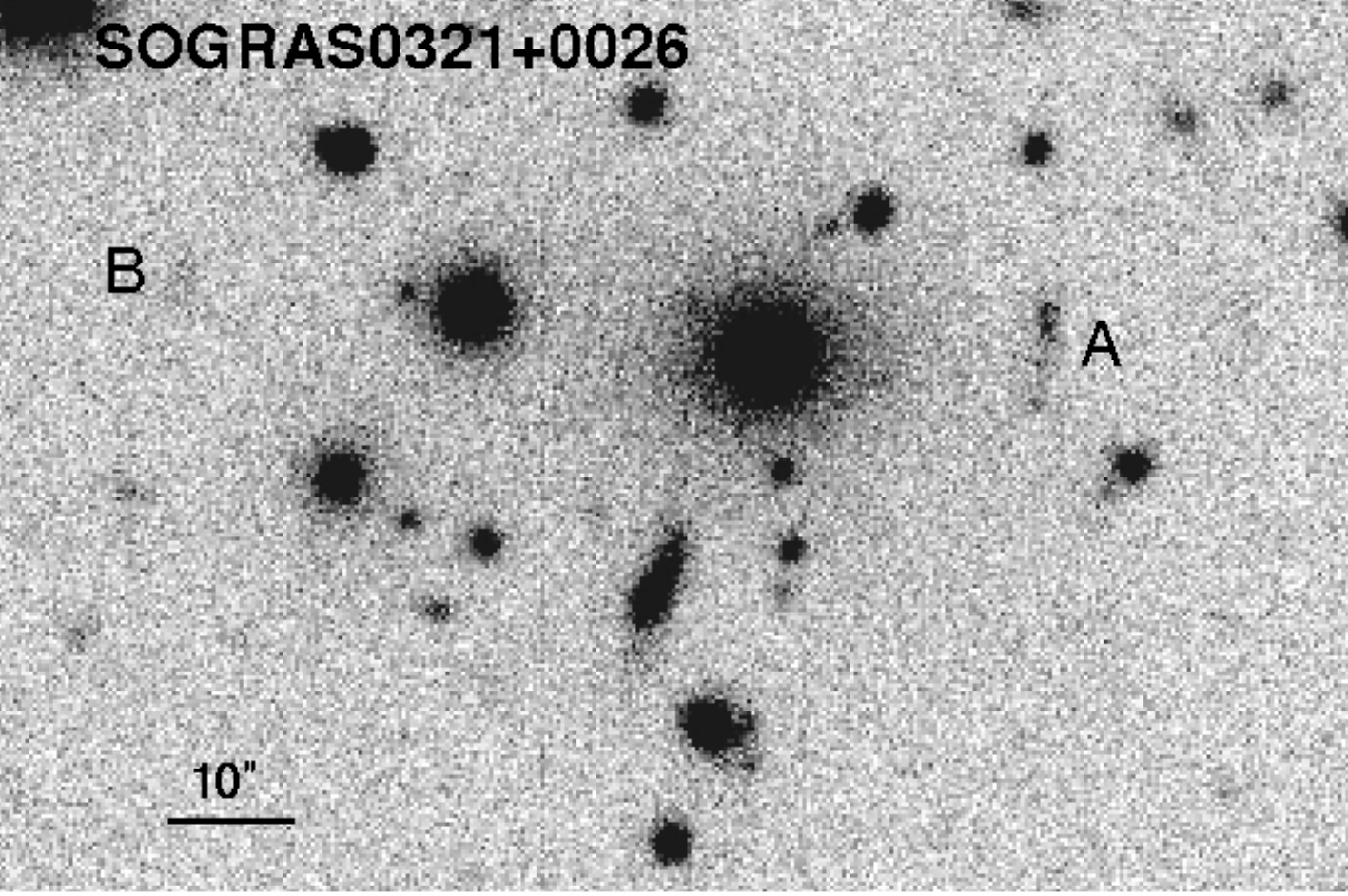}
\includegraphics[scale=0.36]{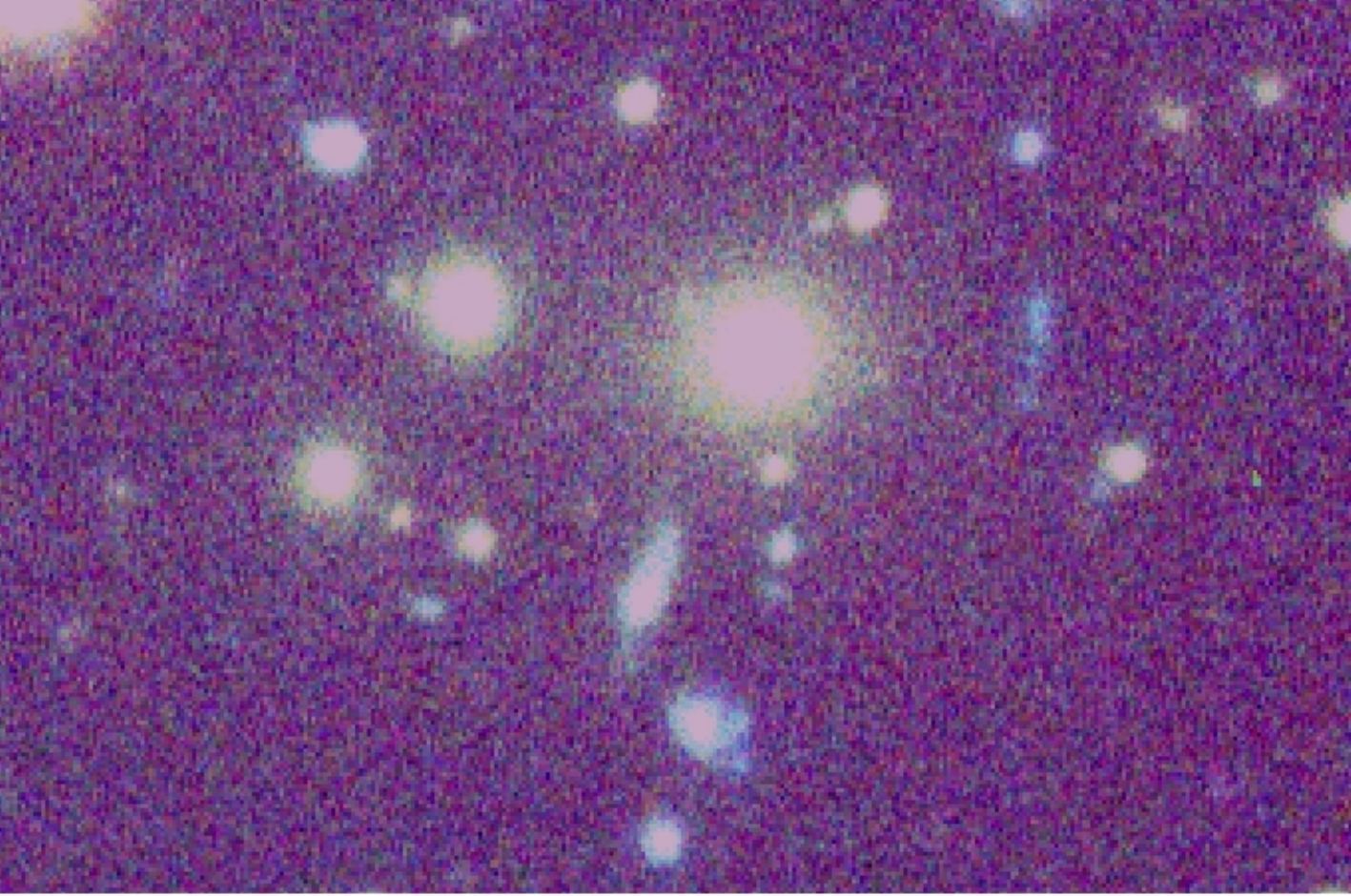}
\includegraphics[scale=0.36]{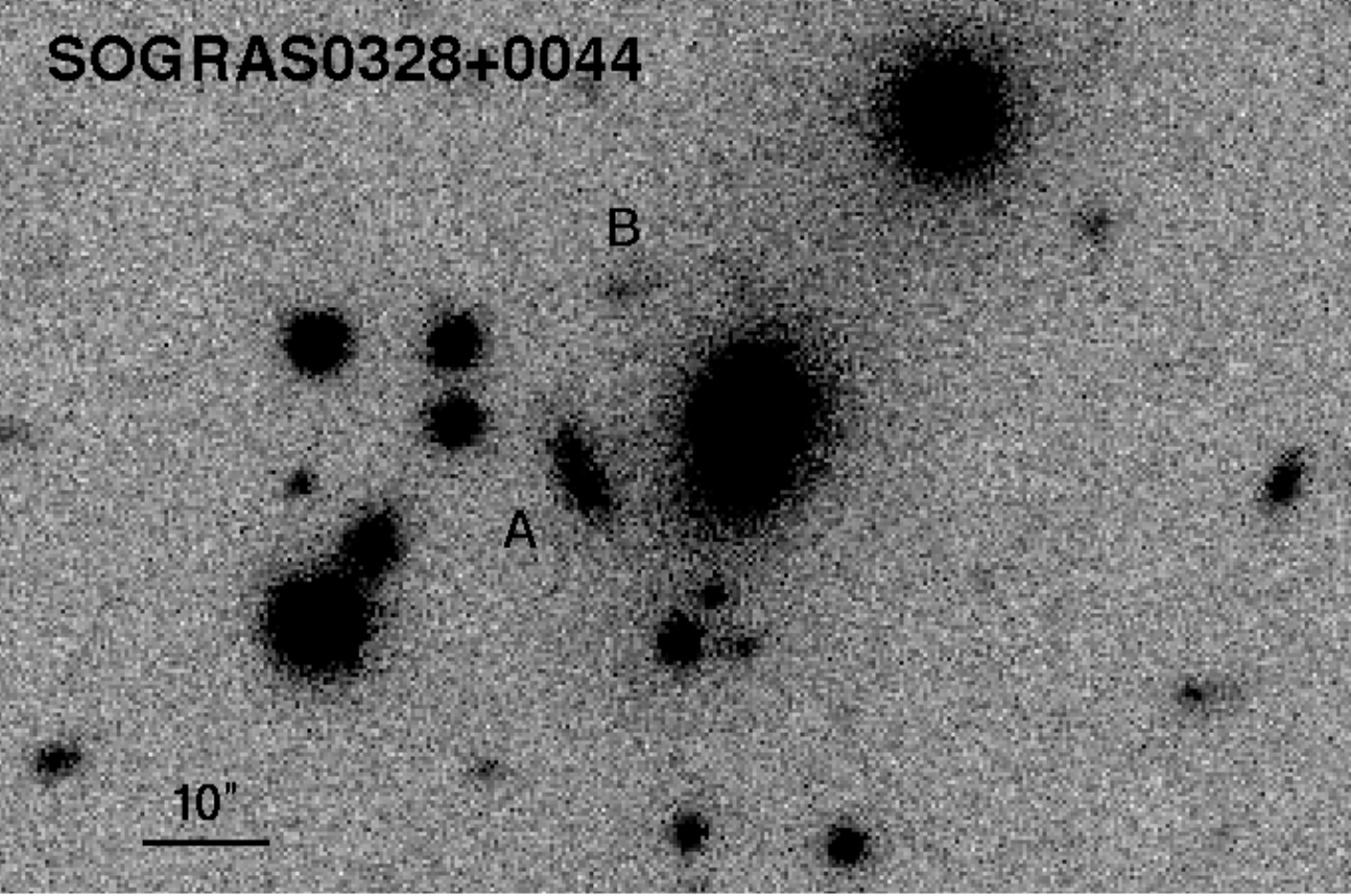}
\includegraphics[scale=0.36]{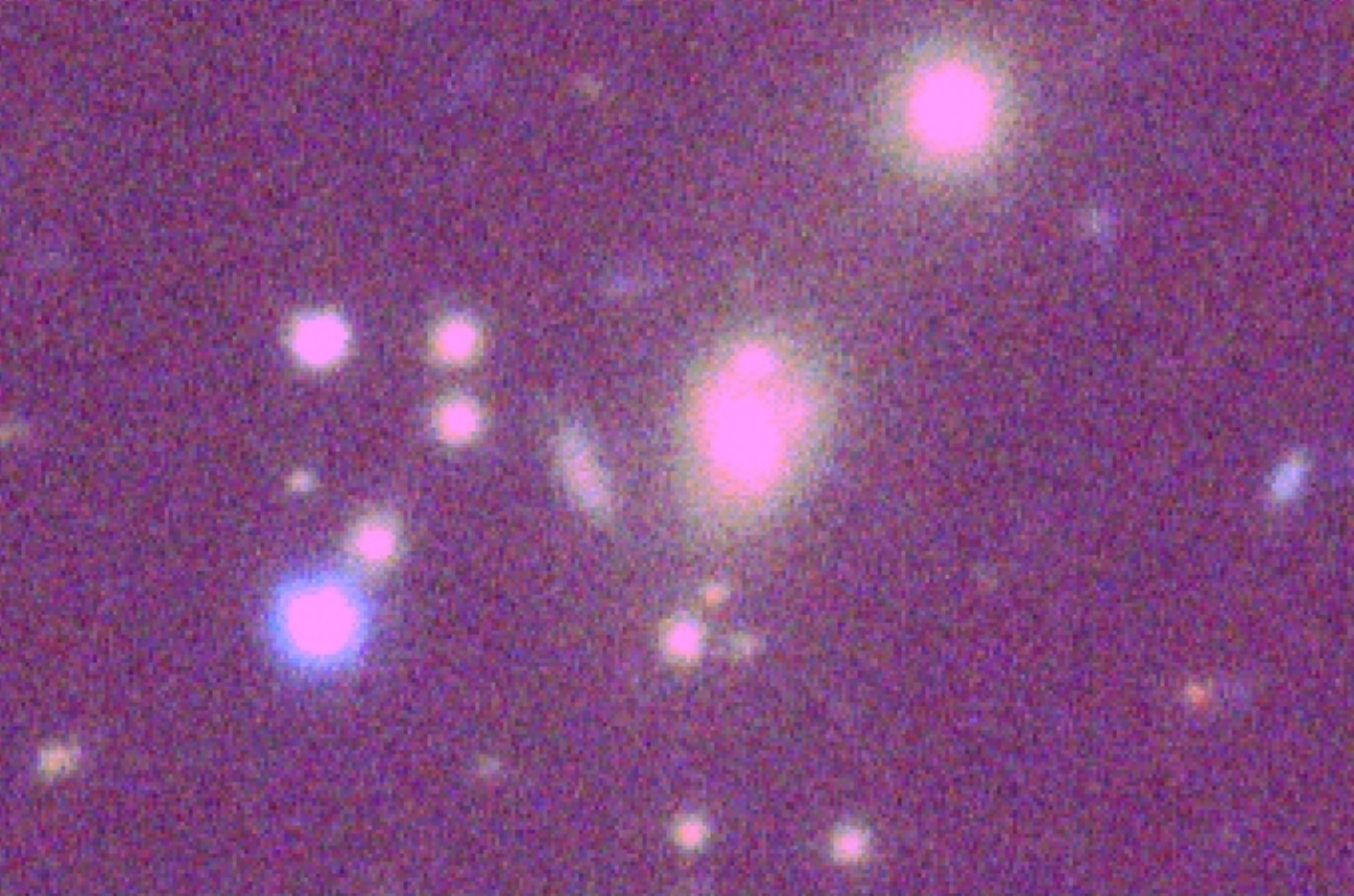}
\includegraphics[scale=0.36]{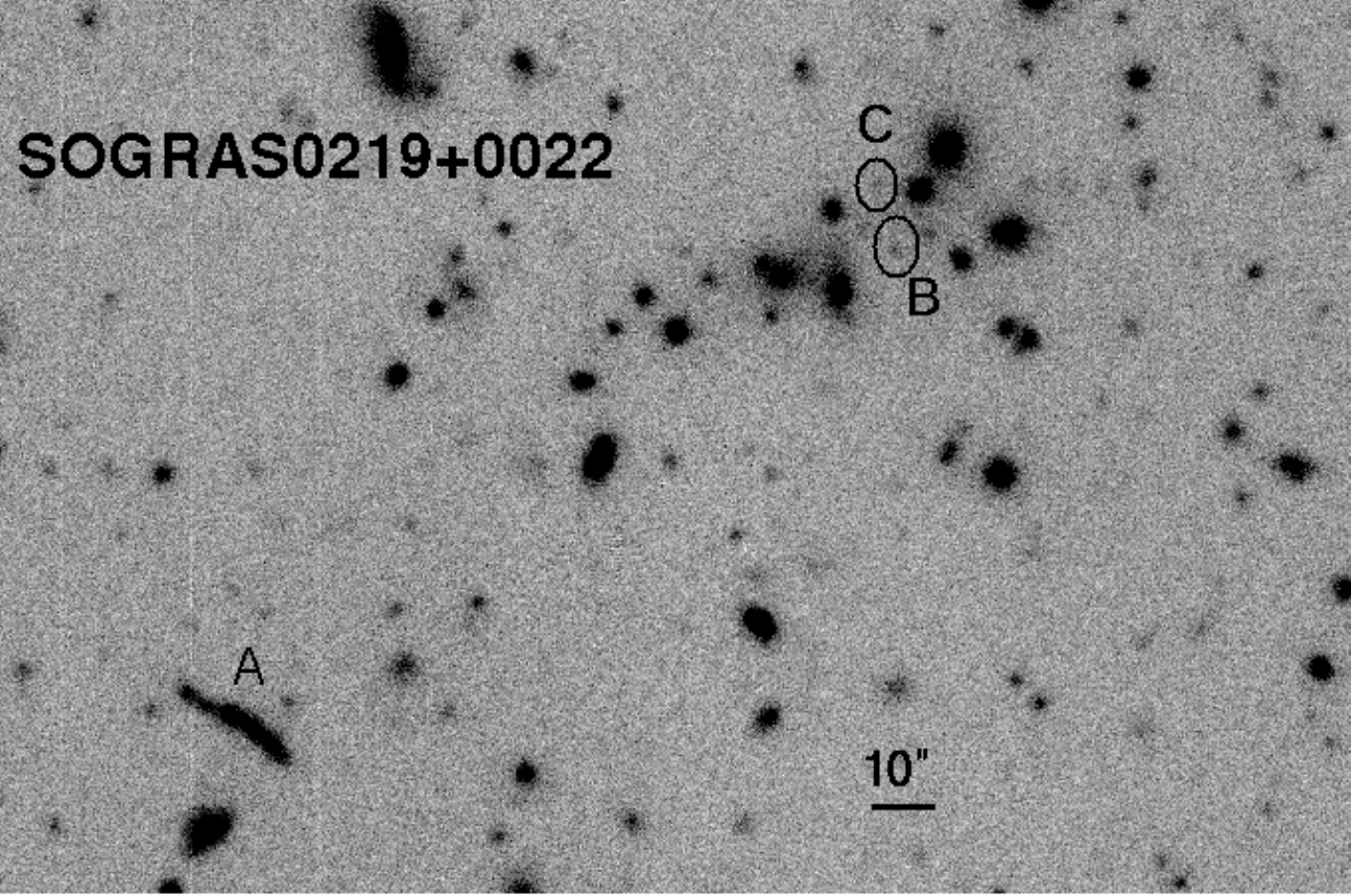}
\includegraphics[scale=0.36]{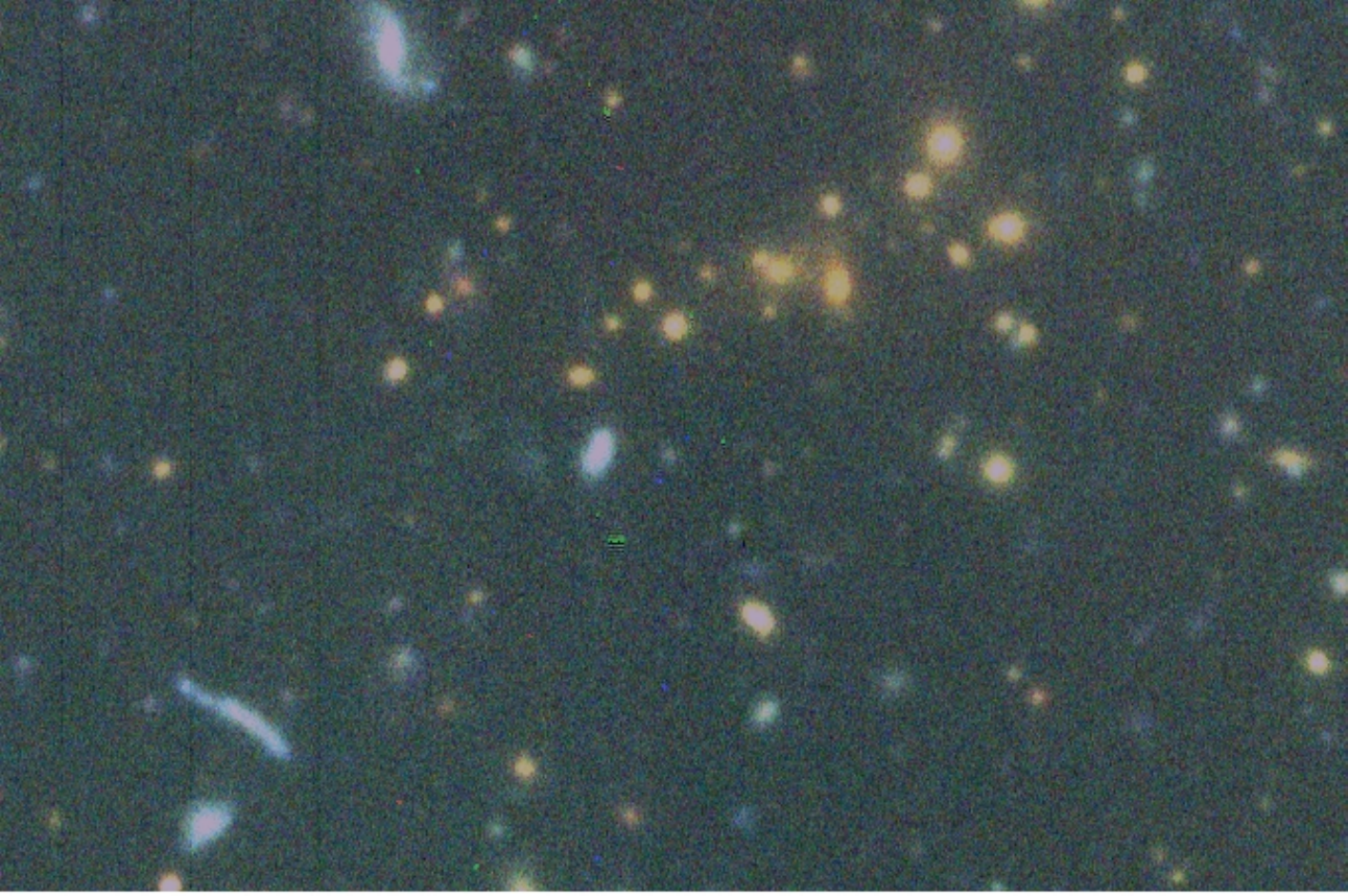}
\includegraphics[scale=0.36]{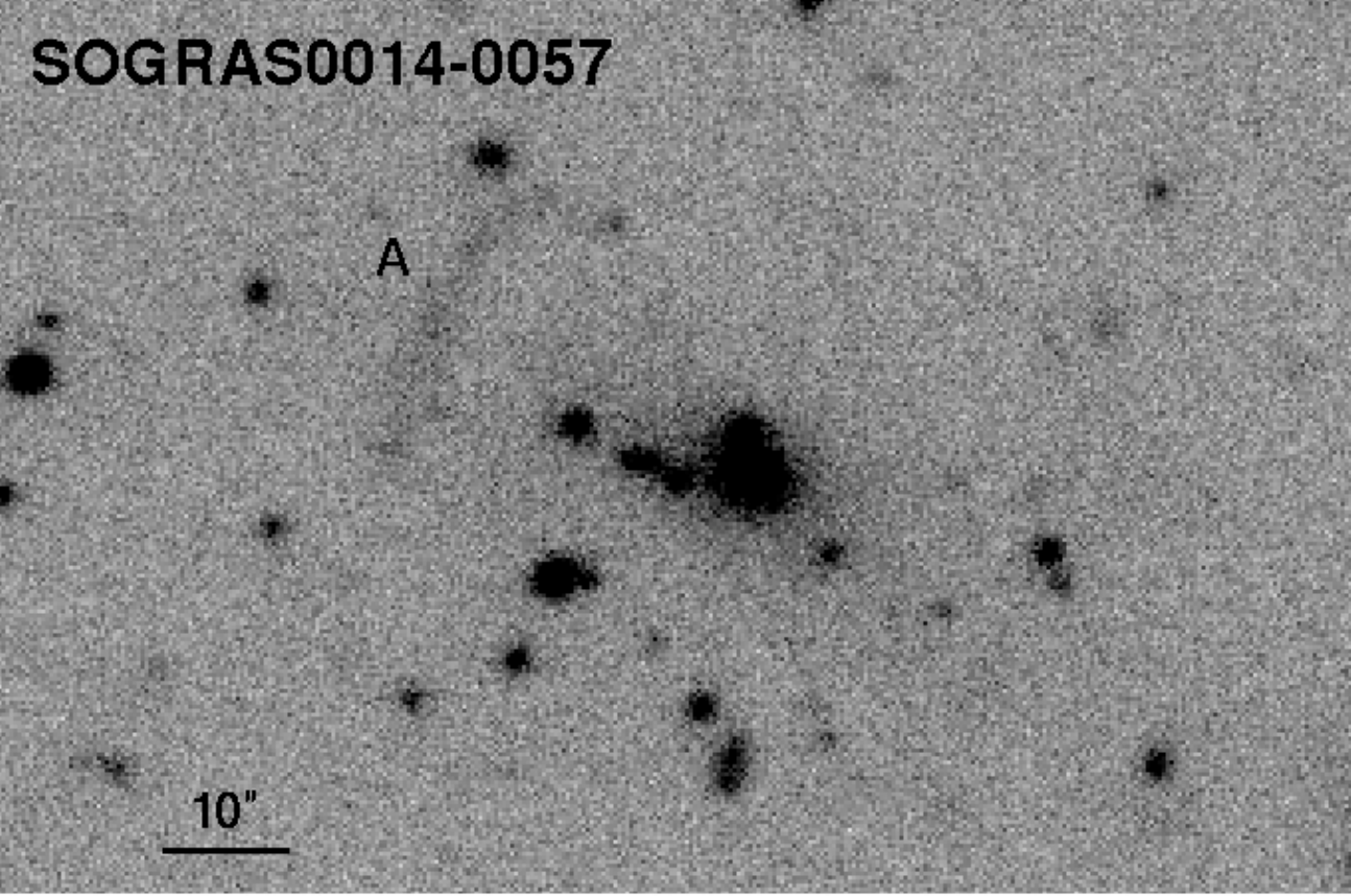}
\includegraphics[scale=0.36]{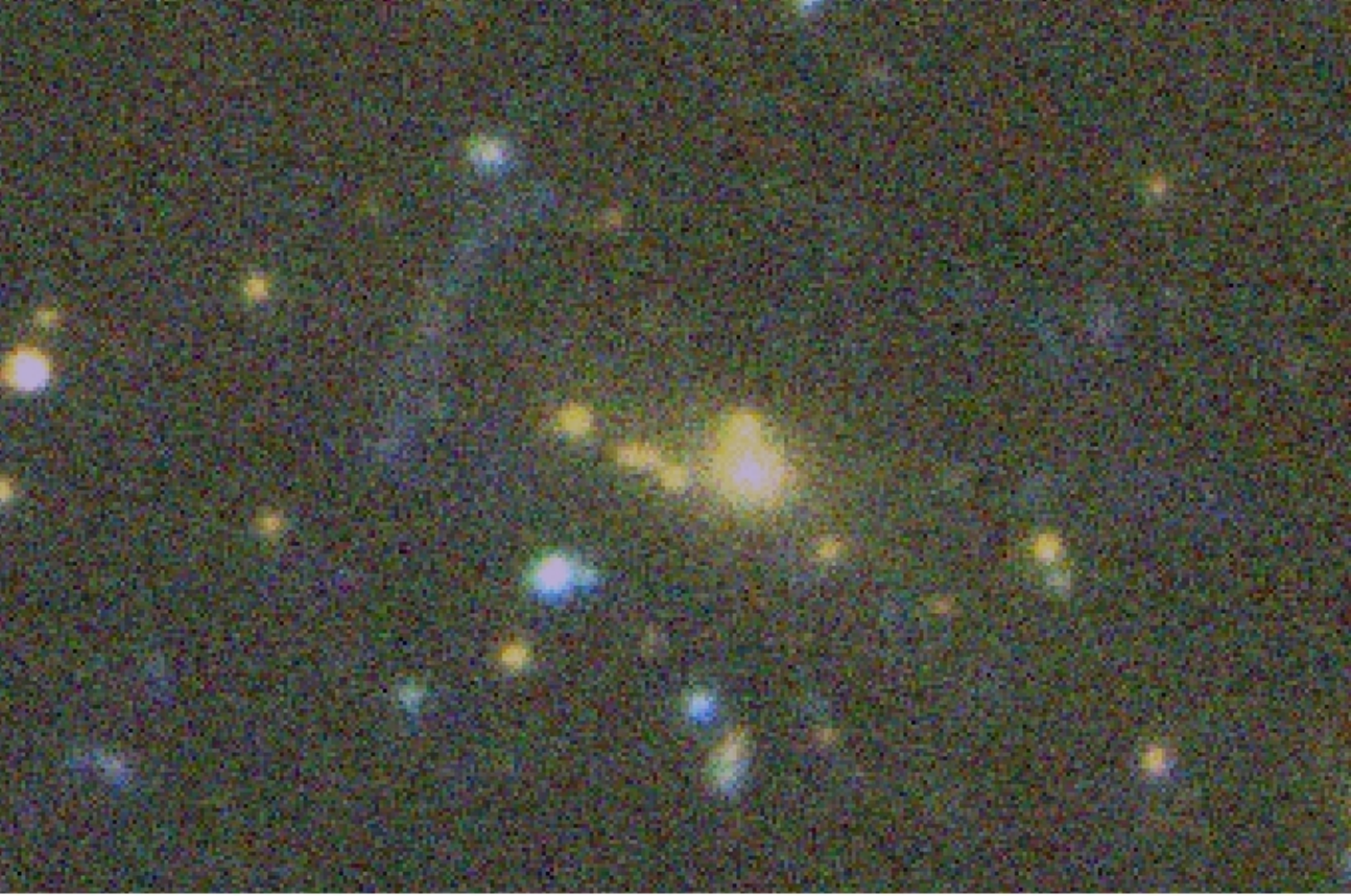}
\caption{(colour online) Strong lensing candidates identified in SOGRAS data. 
In all cases, the left panel is the $g+r+i$ coadded image 
and the right panel is the colour-composite image. }
\label{arc_candidates}
\end{minipage}
\end{figure*}
\addtocounter{figure}{-1}
\begin{figure*}
\begin{minipage}[b]{0.98\linewidth}
\includegraphics[scale=0.36]{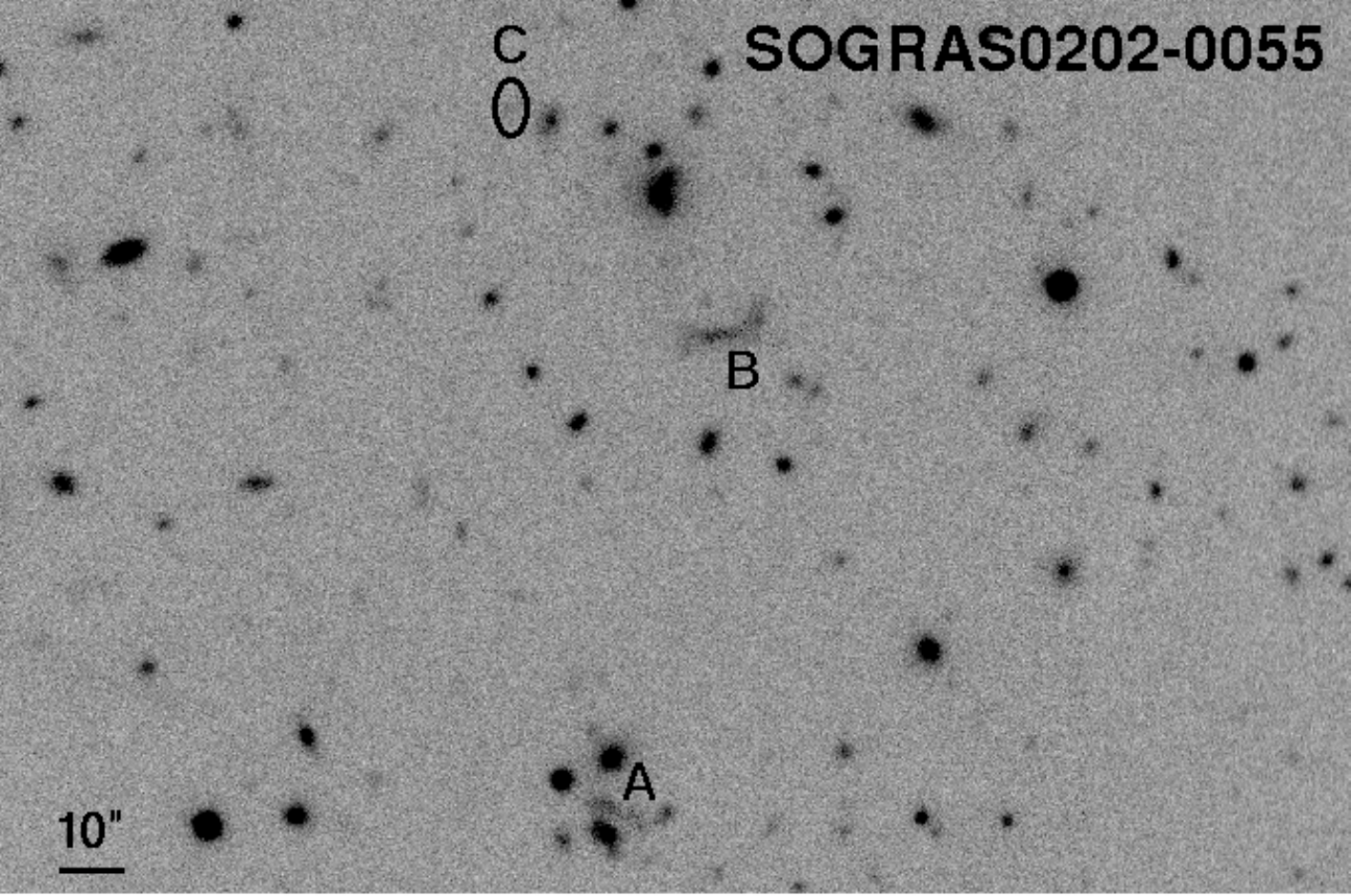}
\includegraphics[scale=0.36]{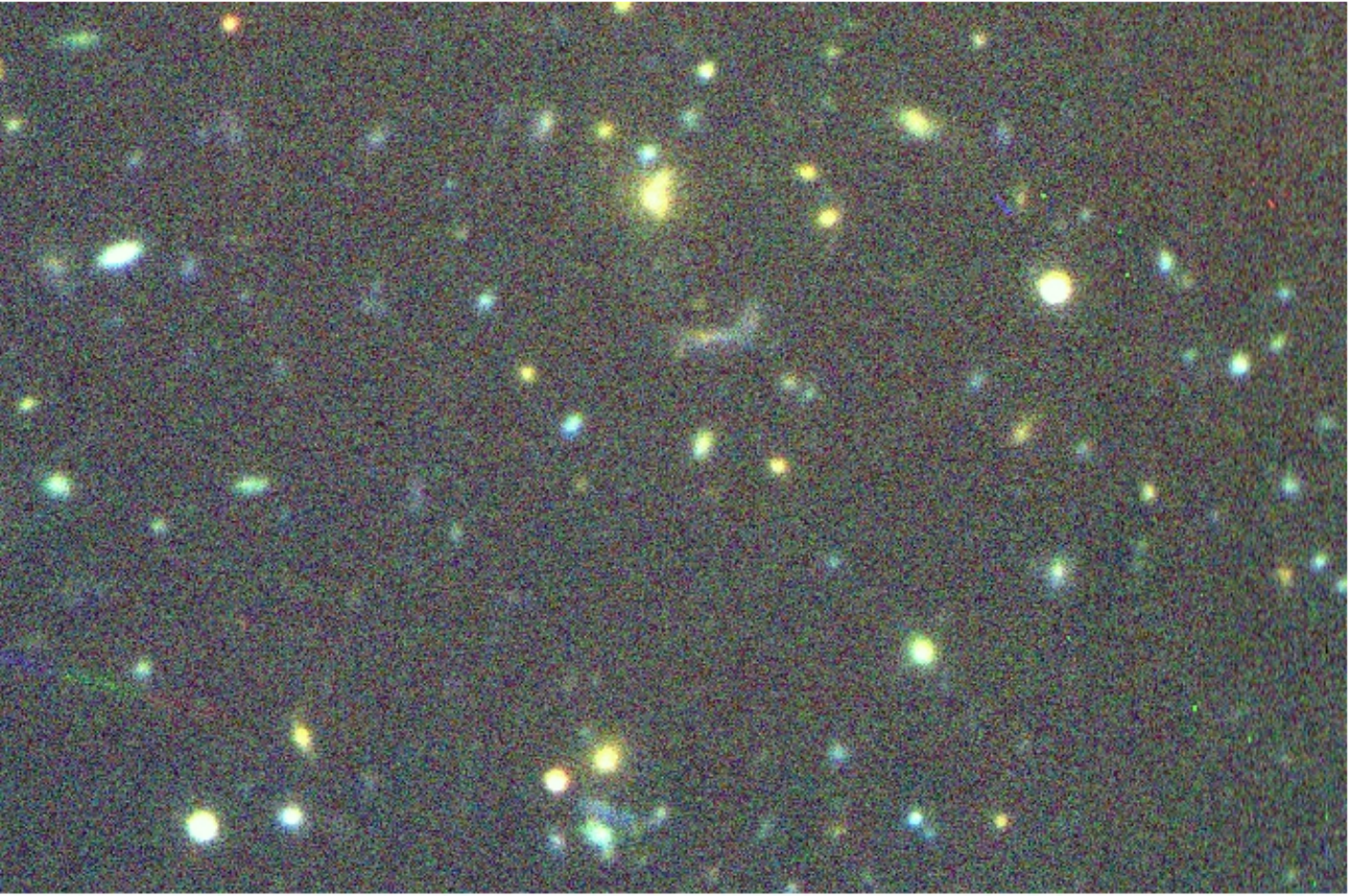}
\includegraphics[scale=0.36]{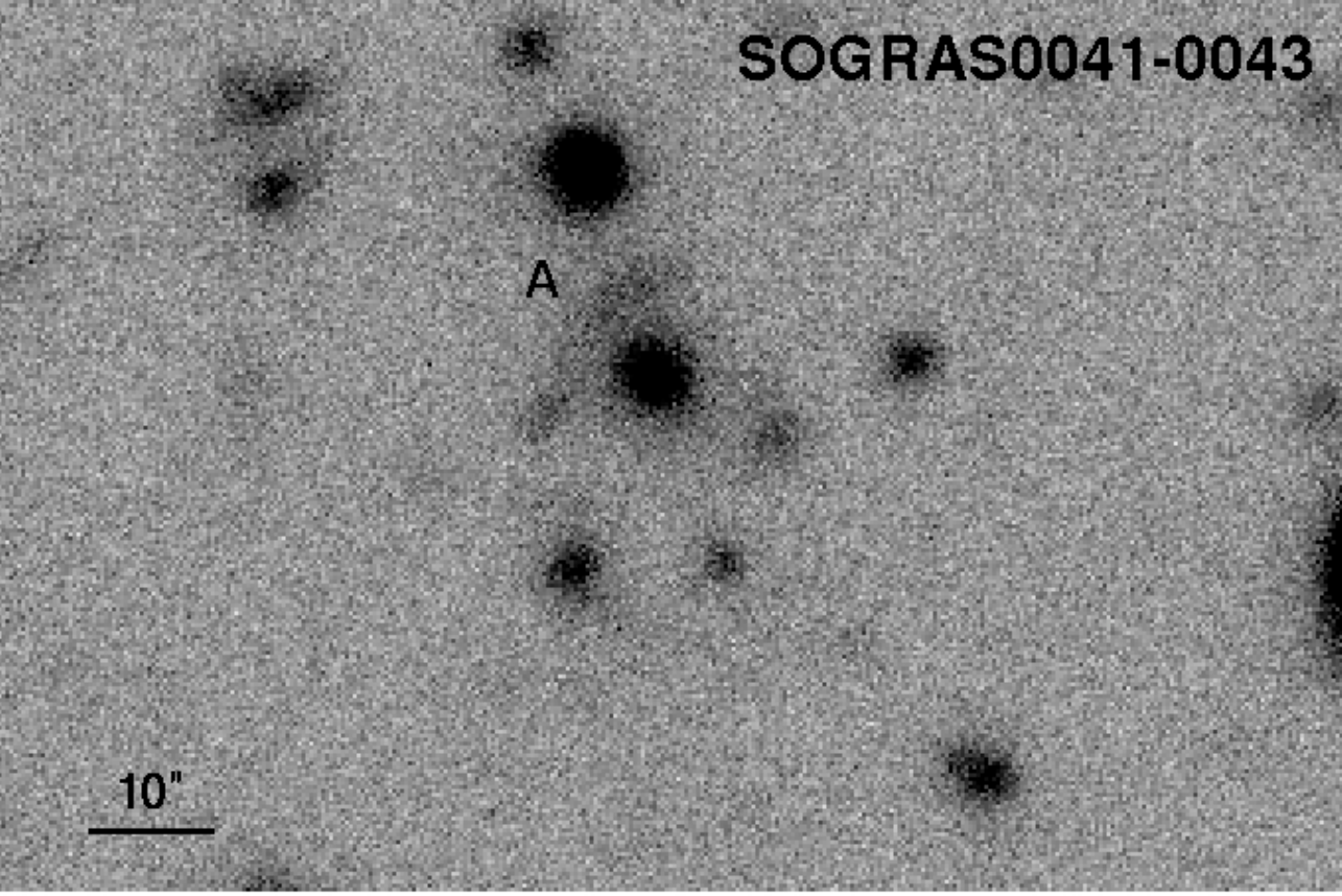}
\includegraphics[scale=0.36]{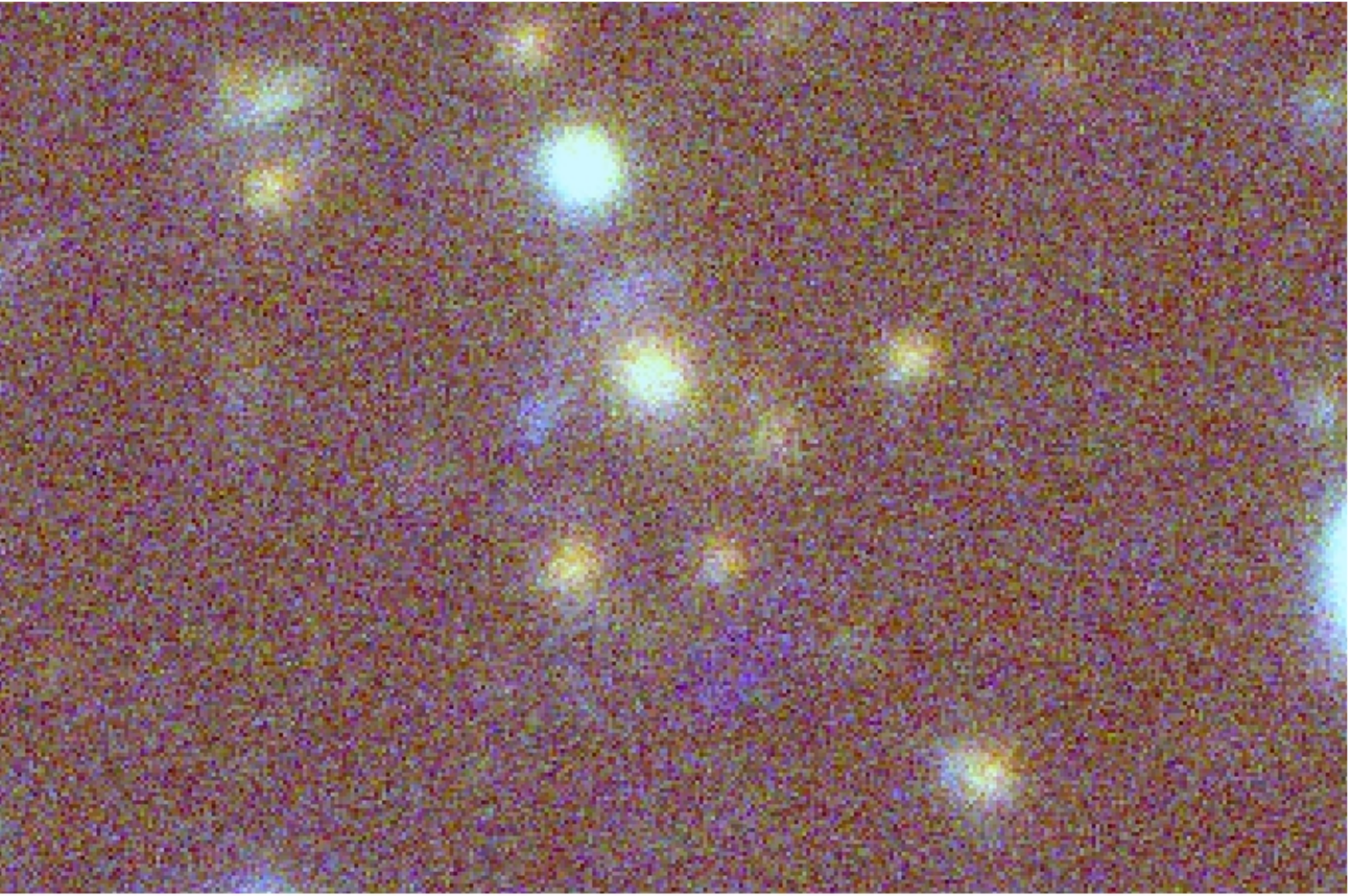}
\includegraphics[scale=0.36]{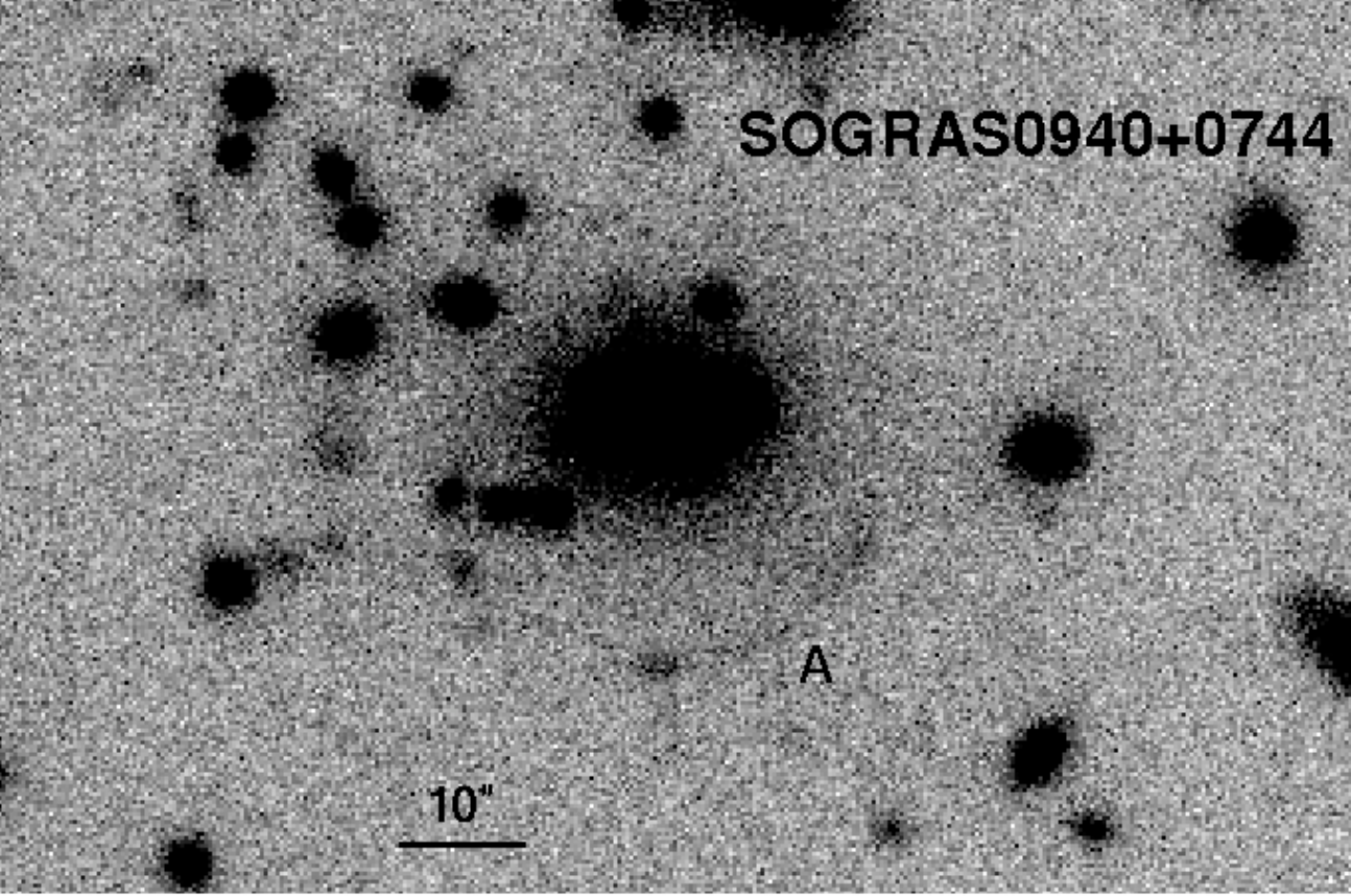}
\includegraphics[scale=0.36]{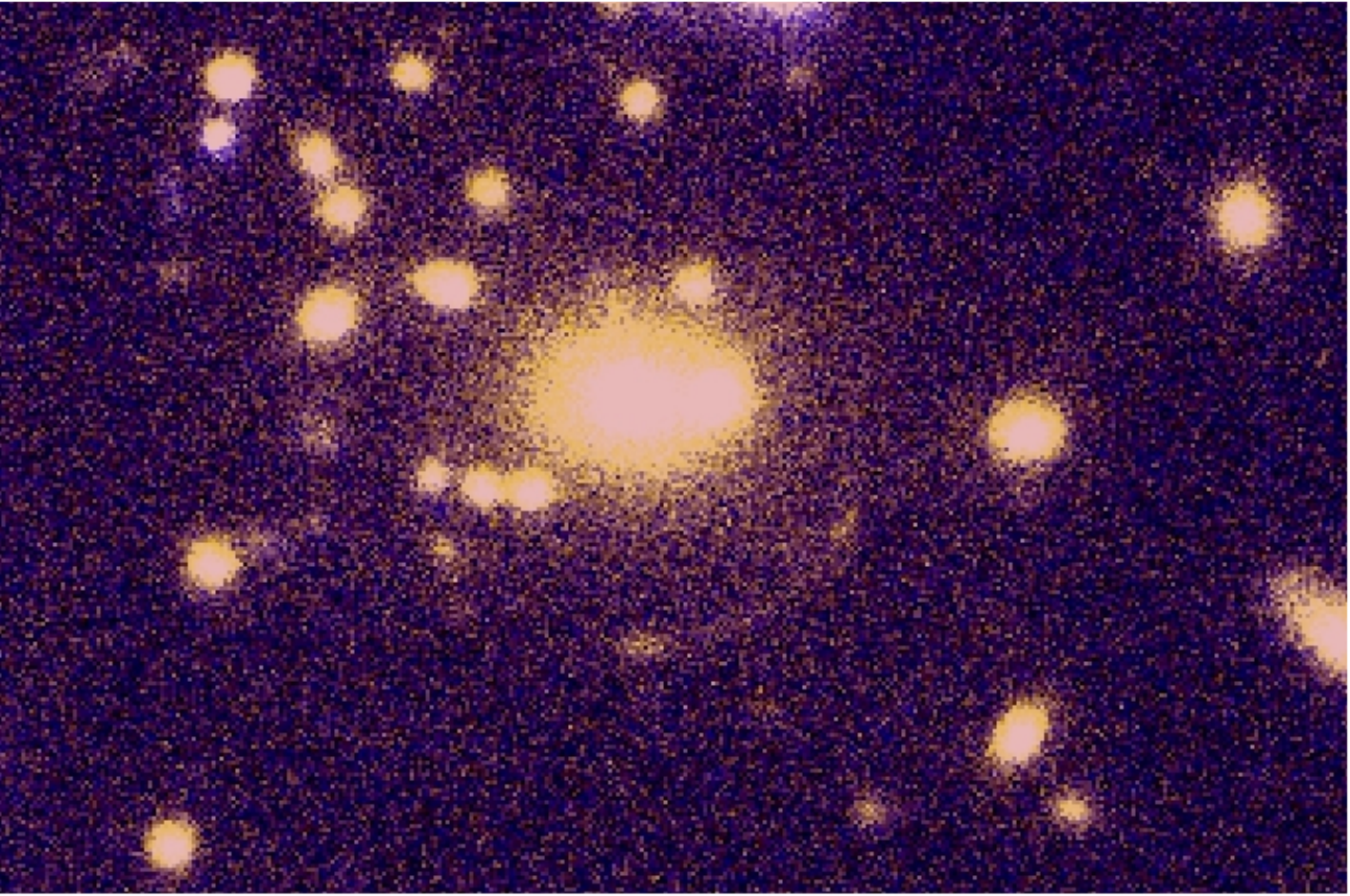}
\includegraphics[scale=0.36]{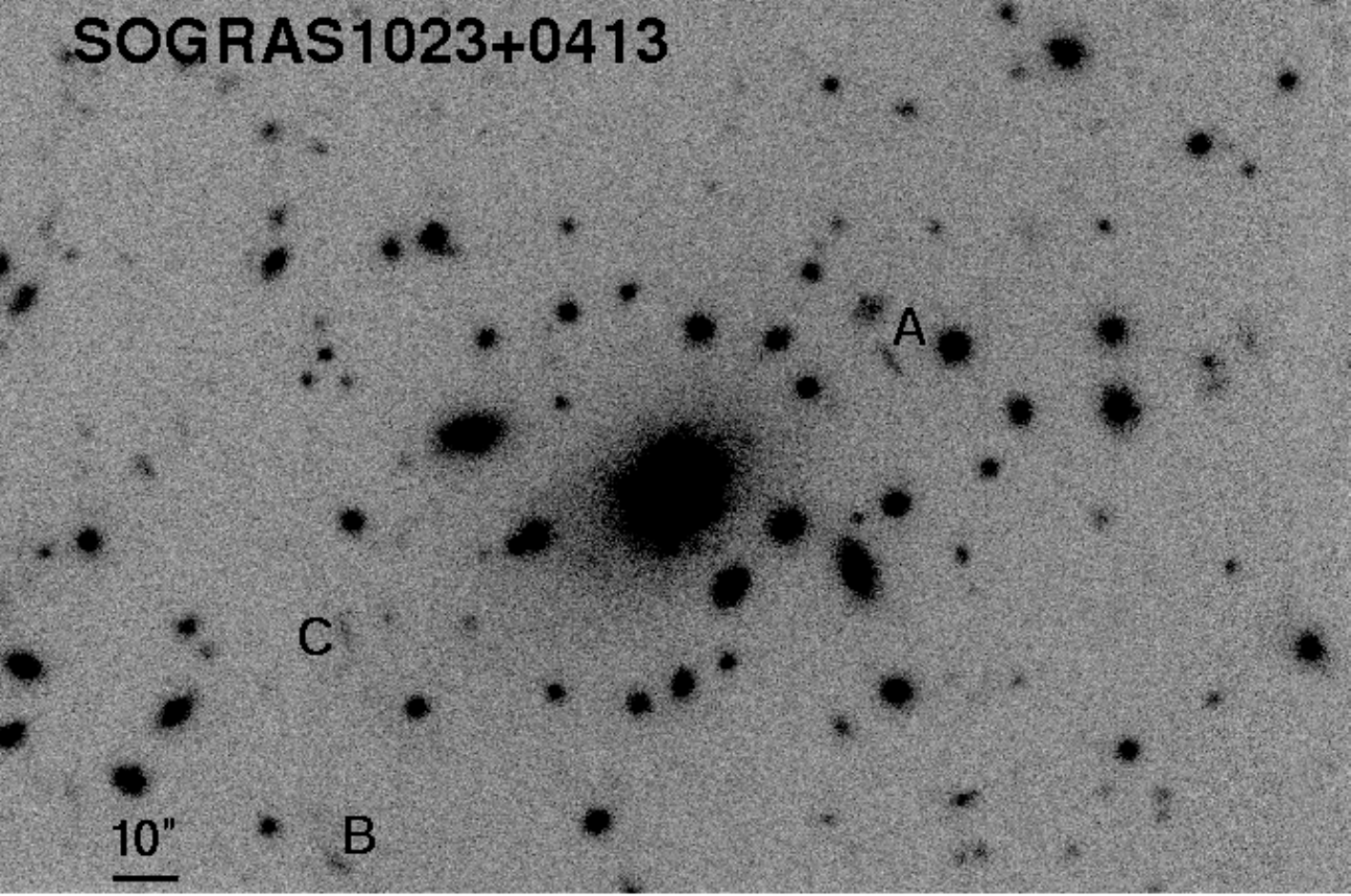}
\includegraphics[scale=0.36]{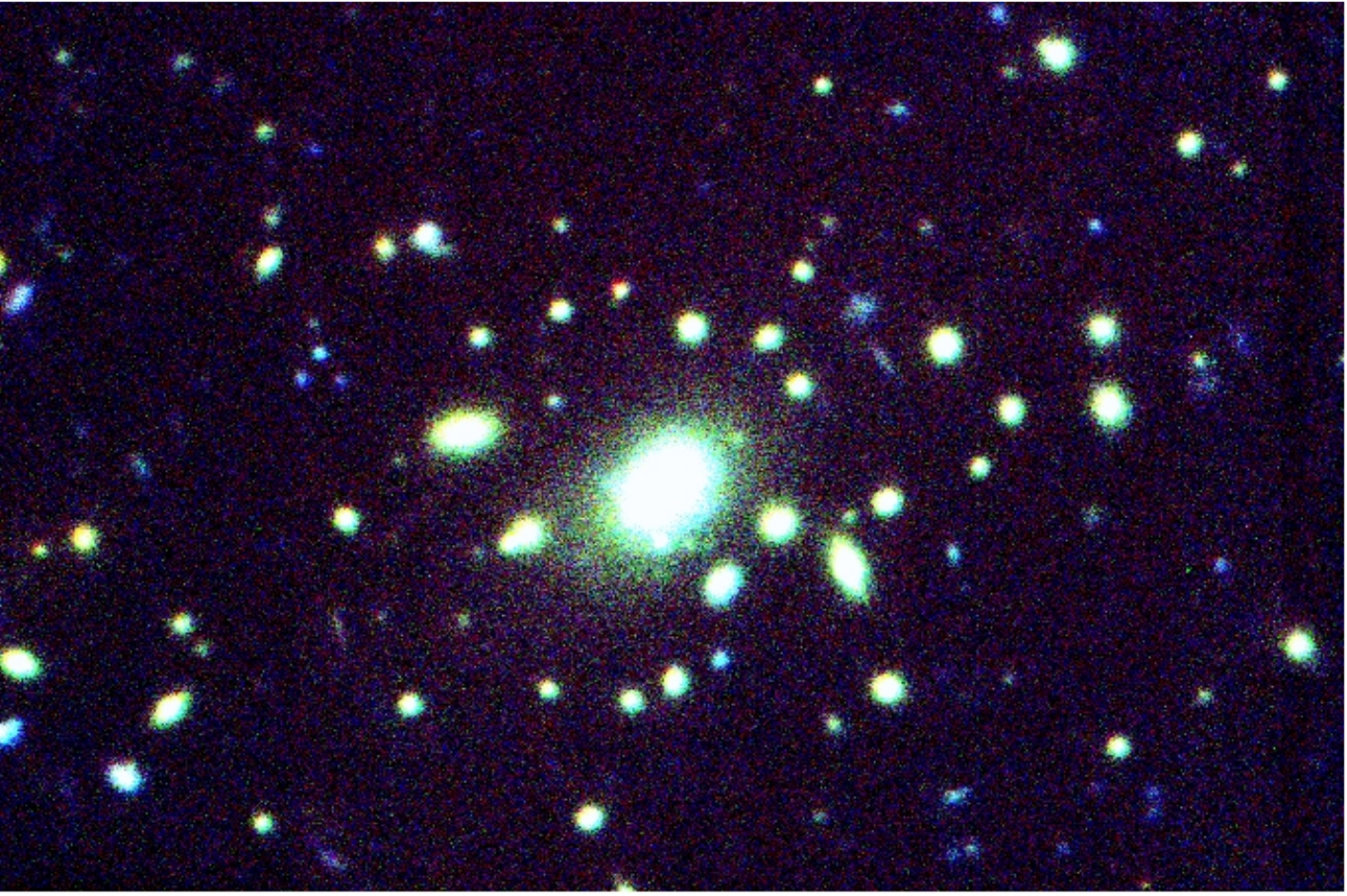}
\caption{$-${\it Continued}}
\label{arc_candidates}
\end{minipage}
\end{figure*}

\begin{table*}
\begin{center}
\caption{Properties of the arc candidates identified in SOGRAS data. 
The photometric redshift $z_{phot}$ of the clusters was taken from Table 1. }
\label{tab_arcs}
\begin{tabular}{|c|c|c|c|c|c|c|c|c|c|c|} \cline{1-11}
Cluster ID         & $z_{phot}$ & Arc ID           &   RA              &  Dec              & $g$   & $r$    &  $i$    & $L$   & $W$           & $L/W$ \\
                  &             &                   &(J2000) & (J2000)                      &    &       &      &(arcsec) &(arcsec) & \\
 \hline 
\multicolumn{1}{|c|}{\multirow{2}{*}{SOGRAS0321+0026}}     &\multicolumn{1}{|c|}{\multirow{2}{*}{0.309}}        &  A  &  03:21:10.55       & 00:26:20.64        & 22.84     & 22.41     & 22.61     & 5.08  &       1.08    & 4.71 \\ \cline{3-11}
\multicolumn{1}{|c|}{\multirow{2}{*}{ }}                & \multicolumn{1}{|c|}{\multirow{2}{*}{ }}  & B &  03:21:12.76        & 00:26:23.73        & 24.13     & 23.63     & 22.90     & 1.23  &       0.62    & 2.00 \\ \hline  
\multicolumn{1}{|c|}{\multirow{2}{*}{SOGRAS0328+0044}}    & \multicolumn{1}{|c|}{\multirow{2}{*}{0.322}} & A & 03:28:15.79        & 00:44:49.59        & 22.73     & 20.72     & 20.25     & 5.93  &       3.03    & 1.95 \\  \cline{3-11}
\multicolumn{1}{|c|}{\multirow{2}{*}{ }}                & \multicolumn{1}{|c|}{\multirow{2}{*}{ }}&B & 03:28:15.34        & 00:44:57.12        & 24.29     & 21.84     & 21.39     & 2.70  &       1.39    & 1.94 \\ \hline 
\multicolumn{1}{|c|}{\multirow{1}{*}{SOGRAS0014-0057}}    & \multicolumn{1}{|c|}{\multirow{1}{*}{0.535}} & A& 00:14:54.93        & -00:57:02.44       & 21.84     & 21.21     & 20.65     & 12.24  &       1.31    & 9.34  \\ \hline   
\multicolumn{1}{|c|}{\multirow{1}{*}{SOGRAS0041-0043}}     &\multicolumn{1}{|c|}{\multirow{1}{*}{0.564}}        &  A  &  00:41:09.22       & -00:43:47.38        & 22.65     & 21.57     & 20.89     & 9.16  &       1.40    & 6.54 \\  \hline  
\multicolumn{1}{|c|}{\multirow{1}{*}{SOGRAS0940+0744}}     &\multicolumn{1}{|c|}{\multirow{1}{*}{0.390}}        &  A  &  09:40:53.33       & 07:44:17.48        & 23.52     & 21.53     & 21.95     & 11.70 &       1.23    & 9.51 \\ \hline  
\multicolumn{1}{|c|}{\multirow{3}{*}{SOGRAS1023+0413}}     &\multicolumn{1}{|c|}{\multirow{3}{*}{0.465}}        &  A  &  10:23:38.59     & 04:11:20.77        & 24.43     & 23.17     & 22.41     & 2.77  &       0.8 0   & 3.46 \\ \cline{3-11}
\multicolumn{1}{|c|}{\multirow{3}{*}{ }}                & \multicolumn{1}{|c|}{\multirow{3}{*}{ }}  & B &   10:23:41.45        & 04:10:41.62       & 24.36     & 23.19    & 22.57     & 2.93  &       1.10    & 2.66 \\  \cline{3-11}
\multicolumn{1}{|c|}{\multirow{3}{*}{ }}                & \multicolumn{1}{|c|}{\multirow{3}{*}{ }}  & C &   10:23:41.43      &   04:11:00.11      & 23.91     & 23.44     & 22.71     & 2.39  &       0.93    & 2.57 \\ \hline  
\end{tabular}
\end{center}
\end{table*}

From the 8 lens system candidates, 2 of them are in the low-$z$ bin, 4 of them are in the high-$z$ bin and the remaining two are in the extra sample.
Concentrating on the 6 lensing systems from the low and high $z$ samples, 
we infer that about 10\% of the clusters have arcs around them. This overall
efficiency is in agreement with larger arc surveys, such as \citet{Gladders03} and \citet{Hennawi2008}. 
Despite the low number statistics, the results are also in qualitative
agreement with the models in that they predict the high-$z$ bin to have
a larger efficiency in arc formation (e.g., Caminha et al., in preparation). 

Follow-up observations of the first 3 lens systems candidates identified in SOGRAS images (in clusters SOGRAS0321+0026, SOGRAS0328+0044 and SOGRAS0219+0022) were conducted on the 8 m Gemini Telescope with the Gemini Multi-Object Spectrograph \citep{SOGRASgemini}. The main aims of this follow-up programme were to confirm spectroscopically the gravitational lensing nature of these candidates, provide mass estimates for the clusters from the velocity dispersion of their member galaxies and perform strong lensing reconstruction of the projected mass distribution of the lenses. The results will be shown and discussed in a forthcoming paper.

Besides the multi-object spectroscopy, we obtained deep imaging to search for new arcs and to determine properties of the lensed galaxies (sources) such as their stellar populations and star formation rate.  
Visual inspection on these deeper images confirms the arc candidates found in clusters SOGRAS0321+0026 and SOGRAS0328+0044, but revealed that the candidates of the cluster SOGRAS0219+0022 are unlikely to be arcs.
The two of them which lie closer to the cluster are clearly seen as point
sources in these images, while the third  
revealed itself as a superposition of two relatively edge-on galaxies.
On the other hand, from a visual inspection of these images, several arc candidates are found in all of them, including in SOGRAS0219+0022.

We thus conclude from our preliminary analysis of the deeper Gemini data that
the number of lensing clusters has not been changed relative to our SOAR
based search, although some individual arc candidates have been added and some
removed. 

\section{Summary and Future Perspectives}
\label{conclusion}

We presented the first results from SOGRAS, an imaging survey towards
47 galaxy clusters using the SOAR telescope. We carefully assessed the quality
of our data products. We estimate our galaxy detection limits as $g \simeq
23.5$, $r \simeq
23$ and $i \simeq 22.5$ at S/N~$\simeq 3$. Photometric calibration was performed
using the SDSS stars in common with our fields with systematic uncertainties
amounting to $0.002$, $0.006$ and $0.005$ mag in $g$, $r$ and $i$ respectively, and
with essentially no systematics. Galaxy photometry suffered from a systematic
offset typically of $0.04-0.12$ mag in comparison to SDSS, likely caused
by the different measurement methods used. Our source catalogue has over 19,000
entries, about 90\% of which are galaxies. We confirm that the data are of enough quality to allow a clear red sequence to be seen in most clusters
in the low-$z$ bin. Furthermore, seeing effects have not strongly affected
shape measurements from our images, as attested by the fact that the
distribution of axis ratios from our data closely resembles that for
nearby cluster galaxies.

Although the number of targets in SOGRAS is smaller than in other arc surveys, the strength of our survey resides in the focus on two narrow redshift intervals whose differential lensing efficiency may 
yield direct information about the evolution of arc incidence. 
This works in a complementary
way to previous studies of arcs around galaxy clusters, some of which have a larger overall statistics in the complete survey, but not in these redshift bins \citep[e.g.,][]{Hennawi2008}. The observations where carried out in similar conditions, with very good seeing, and with the same instrument, assuring an homogeneity of the data.
Furthermore, our sample is unbiased, in the sense that there was no selection based on an {\it a priori} likelihood for an individual cluster to have arcs. These factors make the SOGRAS sample well suited for arc statistics studies, despite the relatively low number of objects.

Preliminary results from a visual inspection suggest an overall efficiency for arcs of about 10\%, consistent with previous studies 
\citep{Hennawi2008,Gladders03}. 
A detailed study of arcs in SOGRAS, including 
comparison with model predictions, mass estimates from arcs, quantitative studies on arc morphology and arc detection will be presented in a separate paper.

Besides the strong lensing studies, this good image quality data can be used to perform a high signal-to-noise weak-lensing analysis by stacking the weak lensing signal of all clusters in a given redshift bin to obtain an overall mass estimate for the clusters. We will also use the arcs and other strong
lensing features to constrain the individual masses of the clusters \citep[e.g.,][]{Cypriano2005}.
The data will also be used for detailed galaxy morphological studies using model-fitting methods, including the modeling of the PSF.

As far as we know, this is the first arc survey that specifically targeted optically selected clusters from the deep Stripe 82 {\it coadd}. This enabled us to select our low-$z$ and high-$z$ samples.
Furthermore, this opens the possibility to exploit the combination with other surveys, by cross matching with the wealth of data in that region of the sky. In particular, most fields are in the CS82 region, which will provide complementary information around the clusters on larger scales.

This dataset will also be used to validate and benchmark arc identification and characterization tools being
developed by our group, including methods to enhance their detectability and arc finding algorithms. 
In some aspects, this small survey can  be seen as a pathfinder for stage III photometric surveys such as DES, given that the SOAR images have similar depth as expected for those surveys.

SOGRAS can also be seen as a feasibility study for a unique arc survey profiting from the adaptive optics capabilities of SOAR in the optical and NIR, which can improve the PSF by a factor of 2 to 5.

\section*{Acknowledgments}

We thank the support of the Laborat\'orio Interinstitucional de e-Astronomia (LIneA) operated jointly by the Centro Brasileiro de Pesquisas F\'isicas (CBPF), the Laborat\'orio Nacional de Computa\c c\~ao Cient\'ifica (LNCC) and the Observat\'orio Nacional (ON) and funded by the Ministry of Science, Technology and Innovation (MCTI). The Brazilian authors of this work are supported by grants from the Conselho Nacional de Desenvolvimento Cient\'ifico e Tecnol\'ogico (CNPq) and Coordena\c{c}\~ao de Aperfei\c{c}oamento de Pessoal de N\'ivel Superior (CAPES). MM is also partially supported by FAPERJ (grant E-26/110.516/2012). ESC also acknowledges support from FAPESP (programme number 2009/07154-8-0).

This paper made extensive use of the database and tools provided by the Sloan Digital Sky Survey (SDSS), including the {\it skyerver} and the {\it catalogue Archive Server}. Funding for SDSS-III has been provided by the Alfred P. Sloan Foundation, the Participating Institutions, the National Science Foundation and the U.S. Department of Energy Office of Science. The SDSS-III web site is http://www.sdss3.org/.
SDSS-III is managed by the Astrophysical Research Consortium for the Participating Institutions of the SDSS-III Collaboration including the University of Arizona, the Brazilian Participation Group, Brookhaven National Laboratory, University of Cambridge, Carnegie Mellon University, University of Florida, the French Participation Group, the German Participation Group, Harvard University, the Instituto de Astrofisica de Canarias, the Michigan State/Notre Dame/JINA Participation Group, Johns Hopkins University, Lawrence Berkeley National Laboratory, Max Planck Institute for Astrophysics, Max Planck Institute for Extraterrestrial Physics, New Mexico State University, New York University, Ohio State University, Pennsylvania State University, University of Portsmouth, Princeton University, the Spanish Participation Group, University of Tokyo, University of Utah, Vanderbilt University, University of Virginia, University of Washington and Yale University.

\end{document}